\documentclass[preprint2]{aastex}

\usepackage[caption=false]{subfig}
\usepackage{graphicx}
\usepackage{epstopdf}
\usepackage{amsmath}
\usepackage{lscape}
\usepackage{natbib}
\usepackage{url}
\usepackage{floatfig}
\usepackage{color}

\usepackage{widetext}

\shorttitle{Robust High-Precision Time-Series Photometry From Well-Sampled Images}
\shortauthors{Chang et al.}

\begin{document}
\captionsetup[subfigure]{labelformat=empty}
\title{A New Method For Robust High-Precision Time-Series Photometry From Well-Sampled Images: Application to Archival MMT/Megacam Observations of the Open Cluster M37}
\author{S.-W. Chang\altaffilmark{1}, Y.-I. Byun\altaffilmark{2}, and J. D. Hartman\altaffilmark{3}}

\altaffiltext{1}{Institute of Earth$\cdot$Atmosphere$\cdot$Astronomy, Yonsei University, Seoul 120-749, South Korea; seowony@galaxy.yonsei.ac.kr}
\altaffiltext{2}{Department of Astronomy and University Observatory, Yonsei University, Seoul 120-749, South Korea; ybyun@yonsei.ac.kr}
\altaffiltext{3}{Department of Astrophysical Sciences, Princeton University, Princeton, NJ 08544, USA}

\begin{abstract}
We introduce new methods for robust high-precision photometry from well-sampled images of a non-crowded field with a strongly varying point-spread function.  For this work, we used archival imaging data of the open cluster M37 taken by MMT 6.5m telescope.  We find that the archival light curves from the original image subtraction procedure exhibit many unusual outliers, and more than 20\% of data get rejected by the simple filtering algorithm adopted by early analysis.  In order to achieve better photometric precisions and also to utilize all available data, the entire imaging database was re-analyzed with our time-series photometry technique (Multi-aperture Indexing Photometry) and a set of sophisticated calibration procedures.  The merit of this approach is as follows: we find an optimal aperture for each star with a maximum signal-to-noise ratio, and also treat peculiar situations where photometry returns misleading information with more optimal photometric index.  We also adopt photometric de-trending based on a hierarchical clustering method, which is a very useful tool in removing systematics from light curves.  Our method removes systematic variations that are shared by light curves of nearby stars, while true variabilities are preserved.  Consequently, our method utilizes nearly 100\% of available data and reduce the rms scatter several times smaller than archival light curves for brighter stars.  This new data set gives a rare opportunity to explore different types of variability of short ($\sim$minutes) and long ($\sim$1 month) time scales in open cluster stars.
\end{abstract}

\section{INTRODUCTION}
We are poised on the threshold of unprecedented technical growth in wide-field time domain astronomy, where ground-based observations yield very precise measurements of stellar brightness from high-volume data streams.  So far, wide-field time-series surveys has been spearheaded by relatively small telescopes since they are supported by large field of view (FOV) instruments operating with high duty cycle (see \citealt{bec04} for a summary of optical variability surveys).  Within the last decade, the advent of large mosaic CCDs has facilitated the coverage of large sky area even for large-aperture telescopes (e.g., MMT Megacam: \citealt{mcl00}; ESO Very Large Telescope Omegacam: \citealt{kui02}; Subaru Suprime-Cam: \citealt{miy02}; CHFT Megacam: \citealt{bou03}; iPTF: \citealt{kul13}).  Although these facilities are generally devoted to imaging surveys, researchers are attempting to utilize them for short- and long-term variability surveys with short-cadence exposures (e.g., \citealt{har08a,pie09,ran11}).  Such wide-field imaging systems have enabled us to observe hundred of thousands of target stars simultaneously and also to detect various variability phenomena.  A remarkable thing about these surveys is that the fraction of variable sources increases as the photometric precision of the survey improves.  For this reason, it is important to improve the accuracy in photometry.
 
Another key issue in wide-field time-series photometry is the removal of temporal systematics from a single image frame or several consecutive image frames.  It has recently become known that systematic trends in time-series data can be different and localized within the image frame when the FOV is large.  Such spatially localized patterns may be related to subtle point spread function (PSF) differences and sky condition within the detector FOV (e.g., \citealt{kov05, pep08, bia09, kim09}).  As these patterns change in time, we can see how the temporal variations of systematic trends affect the brightness and shape of light curves directly.  The time-scale of systematic variation is sometimes comparable to short-term variability, such as transits or eclipses, and in some cases even long-term variability.  Thus, it is often difficult to identify and characterize true variabilities.

In this paper, we introduce a new photometry procedure, called multi-aperture indexing, which is suited to analyzing well-sampled wide-field images of non-crowded fields with a highly varying PSF, such as those produced by wide-field mosaic imagers on large telescopes. We apply this procedure to archival imaging data from the MMT/Megacam transit survey of the open cluster M37 (Hartman et al. 2008a), demonstrating a substantial improvement over the existing photometry. Section 2 describes the MMT imaging database and identifies problems in the existing photometry which motivated the development of our new methods. Section 3 describes the multi-aperture photometry that utilize newly defined contamination index and carefully tuned calibration procedures, including the results of the basic tests to validate our approach.  Section 4 gives an in-depth discussion about systematic trends in time-series data and suggests an efficient way for identifying, measuring, and removing spatio-temporal trends.  Section 5 describes the effects of new calibration on period search, and we summarize our main results in the last section.

\section{ARCHIVAL MMT TIME-SERIES DATA OF M37}
\citet{har08a} have conducted a study to find Neptune-sized planets transiting solar-like stars in the rich open cluster M37.  The observing strategy was carefully designed for a transiting planet search by several considerations (e.g., the reliability of exposure time per frame, the effects of pixel-to-pixel sensitivity variations, and sensitivity of filter).  Their work did not reveal any transiting planets, but it did provide a rare opportunity to explore photometric variability at relatively high temporal resolution with 30--90 s.  \citet{har08b} discovered 1430 new variable stars, including very short-period eclipsing binaries (e.g., V37, V706, V1160) and $\delta$ Sct-type pulsating stars (e.g., V397, V744, V1412).

We used the same data set on the open cluster M37.  A detailed discussion of the observations, original data reduction, and light curve production is described in \citet{har08a,har08b}.  The data archive consists of approximately 5000 $r'$-filter images taken over 24 nights with the wide-field mosaic imager (Megacam) mounted at the $f/5$ Cassegrain focus of the 6.5m MMT telescope.  Note that Megacam is made up of 36 2048 $\times$ 4608 pixel CCD chips in a $9 \times 4$ pattern, covering a 24$\arcmin$$\times$24$\arcmin$ FOV \citep{mcl00}.  This instrument has an unbinned pixel scale of $0^{\prime\prime}.08$, but it was used in $2 \times 2$ binning mode for readout.
 
The observation logs are summarized in Table 1 of \citet{har08a}.  In brief, the $r\rq{}$-band time-series observations were undertaken between 2005 December 21 and 2006 January 21, with a median FWHM of $0.89\pm0.39$ arcsec.  Exposure times are chosen to keep an $r\sim15$ mag star as close to the saturation limit, which is expressed as a function of seeing conditions.  With an average seeing $\sim$0$^{\prime\prime}$.89 on images, the quality of the images is good to achieve high-precision light curves (less than 1\% rms value) down to 20.  In addition to the imaging data set, this database includes light curve data sets for a total of 23,790 sources detected in a co-added reference image.  Theses light curves are obtained by the image subtraction technique using a modified version of ISIS software (\citealt{ala98}; \citealt{ala00}).  

\begin{figure}[t]
\centering
	\includegraphics[width=\linewidth, angle=0]{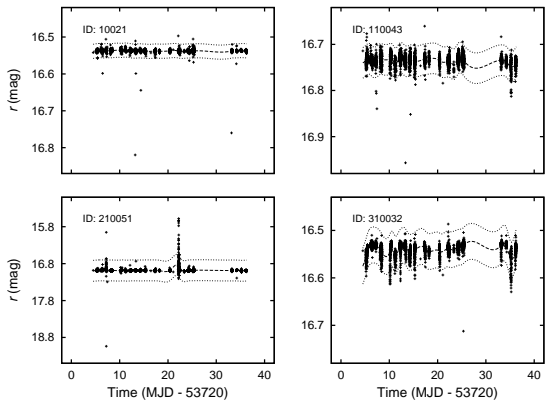}  
	\caption{Example of MMT light curves for the brighter stars in archive.  The dashed lines are weighted spline approximation with $\pm$3-$\sigma$ control limits (dotted lines).  These light curves contain outlier points that significantly increase the rms scatter of the raw light curves.}
	\label{fig:Fig1}
\end{figure}

As shown in Figure \ref{fig:Fig1}, however, the raw light curves from the original image subtraction procedures exhibit many unusual outliers, and more than $\sim$15\% of data get rejected by a simple filtering algorithm after cleaning procedures.  In practice, brutal filtering that is often applied to remove outlying data points can result in the loss of vital data, with seriously negative impact to short-term variations such as flares and deep eclipses.  We also find that the image subtraction technique often resulted in measurement failures from several frames due to poor seeing or tracking problem.  After removing these bad frames, it leads to loss of additional $\sim$5\% data points from most light curves.  In order to overcome this problem, we have re-processed the entire image database with new photometric reduction procedures.

\section{NEW PHOTOMETRIC REDUCTION}
\subsection{Preparation for Photometry}
We followed the standard CCD reduction procedures of the bias correction, overscan trimming, dark correction, and flat-fielding as described in \citet{har08a}. The individual CCD frames were calibrated in IRAF, using the mosaic data reduction package MEGARED.\footnote{The 64-bit version of Megacam reduction package is available from \url{https://www.cfa.harvard.edu/~bmcleod/Megared/.}}  The first step is to correct the pixel-to-pixel zero-point differences that are usually described by the sum of a mean bias level and a bias structure.  As the bias frames were not separately taken during the time of the observations, the mean bias level was subtracted from each image extension using an overscan correction and so we cannot remove any remaining bias structure from all overscan-subtracted data frames.  According to description in Matt Ashby's Megacam reduction guide,\footnote{The detailed reduction procedures of MMT/Megacam are described by Matthew L. N. Ashby at \url{https://www.cfa.harvard.edu/~mashby/megacam/megacam_frames.html}.} 
the bias structure can be very significant in some small regions such as the portions of the arrays close to the readout leads.  The dark currents are normally insignificant for Megacam so that corrections are not needed even for long exposures.  The next step is to correct pixel-to-pixel variations in the sensitivity of the CCD;  we used the program \texttt{domegacamflat2}\footnote{This C routine is written by J. D. Hartman to reduce I/O overhead.  It reads in a list of Megacam mosaic images and a Megacam mosaic flat-field image, and then writes out the flattened image over the existing image.}.  This program determines the scaling factors to correct the gain difference between the two amplifiers of each chip by finding the mode in the quotient of pixels to the left and right of the amplifier boundary, and then flattens each of the frames with a master flat field frame.  It is worth mentioning that the sky conditions were rarely photometric during the observing run, with persistent light cirrus for most of the nights.  Therefore it was only possible to obtain twilight sky flats on a handful of nights (dome flats were not possible).

We removed bad pixels using the Megacam bad pixel masks distributed with the MEGARED package.  The values of bad pixels are replaced with interpolated value of the surrounding pixels using the IRAF task \texttt{fixpix}.  The numerous single pixel events (cosmic-rays) were identified and removed using the LACosmic package \citep{van01}.

The Megacam data already have a rough World Coordinates System (WCS) solution that is based on a single value of the telescope pointing.  To update these with a more precise solution, we applied astrometric correction to each CCD in the mosaic using the WCSTools imwcs program \citep{min02}.  The new solution is derived by minimizing the differences between the R.A. and decl. positions of sources in a single CCD chip and their positions listed in the 2MASS Point Source Catalog \citep{skr06}.  The resulting astrometric accuracy is typically better than 0$\arcsec$.1 rms in both R.A. and decl.

\subsubsection{Building a Master Source Catalog}
Typically, a point or extended source detection algorithm is applied to each frame independently and it always requires criteria for what should be regarded as a true detection.  In obtaining the pixel coordinates for all objects in the M37 fields, this procedure often misses some objects when the detection threshold approaches the noise level.  Also it needs a substantial effort to match the objects that are detected in only some of the frames.  Our approach is as follows: a complete list of all objects is obtained from a co-added reference frame, and then the photometry is performed for each frame using the fixed positions of the sources detected on the reference.  Since the relative centroid positions of all objects are the same for all frames in the time series, we can easily place an aperture on each target and measure the flux even for the stars at the faint magnitude end.

We constructed the reference frame for each chip from the best seeing frames using the SWarp\footnote{SWarp is a program that resamples and co-adds FITS images, distributed by Astrometic.net at \url{http://www.astromatic.net/software/swarp.}} software.  Benefiting from a highly accurate astrometric calibration of input frames, we were able to improve the quality of co-added images.  In the SWarp implementation, the pixels of each frame were resampled using the Lanczos3 convolution kernel, then combined into the reference frame by taking a median or average.  After this was done, sources were detected and extracted on the reference frame using the SExtractor software \citep{ber96}.  When configured with a lower detection threshold, SExtractor extracts the number of spurious detections (e.g., diffraction spikes around bright stars, or outer features of bright galaxies).  These false detections were removed by careful visual inspection for each chip.  The final catalog contains a total of 30,294 objects including both point and extended sources.

\subsubsection{Refining the Centroid of Each Object}
Prior to the photometry, the initial centroid coordinates of the target objects for each frame were computed by using the WCSTools sky2xy routine \citep{min02}.  The stored world coordinate system for each frame is used to convert the (R.A., decl.) coordinates from the master source catalog to the $(x, y)$ pixel locations.  However, actual positions of objects for each frame can be slightly moved from its original locations depending on the focus condition of instrument, seeing condition, and the signal-to-noise ratio (S/N) in the individual observations.  These types of positioning errors (i.e., centroid noise) will lead to the internal error for a photometric measurement that results from placement of the measuring aperture on the object being measured.  The situation gets worse for faint stars because the centroid position of them is itself subject to some uncertainty.  The fractional error in the measured flux as a result of mis-centering is given by:\begin{equation}
\frac{\delta F}{F_\mathrm{0}} \approx \frac{1}{\sqrt{2\pi}}   \frac{\Delta}{\sigma} \frac{2R\Delta}{\sigma^{2}}e^{-R^{2}/2\sigma^{2}}, 
\end{equation} where $\Delta$ is the positioning error, $R$ is the radius of aperture, $\sigma$ is the profile width for a source with a Gaussian PSF, and total flux $F_\mathrm{0}$ (see Appendix A in Irwin et al., 2007 for details).  This expression shows that if we set the aperture radius equal to the FWHM ($R$=2.35-$\sigma$), even small differences in placement of the aperture (e.g., $\Delta$=0.1-$\sigma$) may increase the uncertainty in the flux measurements ($\approx$ 1 mmag).  Thus, accurate centroid determination is important to achieve the high-precision photometry.

Following the windowed centroid procedure in the SExtractor, a refined centroid of each object is calculated iteratively.  On average, the rms uncertainty in the coordinate transformation using the WCS information was $0.044\pm0.01$ pixels for the bright reference stars.  After the coordinate transformation from sky to $xy$, however, the centroid coordinates ($x_\mathrm{ini}$, $y_\mathrm{ini}$) are slightly misaligned from their actual ones ($x_\mathrm{center}$, $y_\mathrm{center}$).  The refined centroid values ($x_\mathrm{final}$, $y_\mathrm{final}$) are used only if the maximum displacement is at least less than 1.5 pixels.  This condition prevents arbitrary shifting of a source centroid, especially for faint stars.

\subsubsection{Estimation of Background Level}
We estimated a local sky background by measuring the mode of the histogram of pixel values within a local annulus around each object, which is suitable choice for our uncrowded field (less than $\sim$1000 stars per chip).  This process is a combination of $\kappa$-$\sigma$ clipping and mode estimation.  The background histogram is clipped iteratively at $\pm$3-$\sigma$ around its median, and then the mode value is taken as:
\begin{equation} 
Mode = 3 \times Median - 2 \times Mean.
\end{equation}
It represents the most probable sky value of a randomly chosen pixel in the sample of sky pixels \citep{ste87}.  For relatively crowded regions, we utilized a background map created by SExtractor package using a mesh of $32\times32$ pixels and a median filter box of $5\times5$ pixels.  This map is used to confirm the properness of individual sky values from annulus estimates.

\subsection{Multi-aperture Indexing Photometry}
Modern data reduction techniques aim to reach photon noise limit and minimize systematic effects.  For example, differential photometry technique can be achieved better than 1\% precision for brighter stars (e.g., \citealt{eve01,har05}), and the deconvolution-based photometry algorithm leads to the minimization of systematic effects in very crowded fields (e.g., \citealt{mag07,gil07}).  However, conventional data reduction methods often fail to handle various artifacts in wide-field survey data.  We present below a new photometric reduction method for precise time-series photometry of non-crowded fields, without the need to involve complicated and CPU intensive process (e.g., PSF fitting or difference image analysis).

\begin{figure}[t]
\centering
  \includegraphics[width=\linewidth, angle=0]{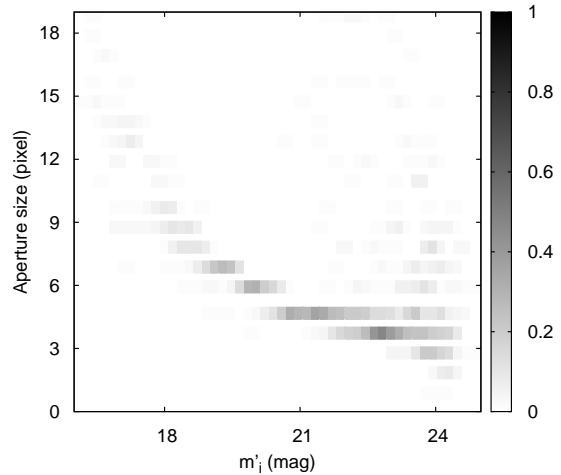}
  \caption{Tendency of optimal aperture selection from the multi-aperture photometry.  As stars get fainter, the optimal aperture decreases in size. Sources outside of this sequence are suspected to be contaminated.}
  \label{fig:Fig2}
\end{figure}

\subsubsection{Photometry with Multiple Apertures}
Our photometry is similar to standard aperture photometry, except in that we compute the flux in a sequence of several apertures and then determine the optimum aperture individually to each object at each epoch.  This multi-aperture photometry is an efficient way to determine the optimum aperture size that gives the maximum S/N for a flux measurement.  The maximum S/N is not necessarily at the same aperture for all objects, and it can be obtained from a relatively small aperture \citep{how89}.  This {\itshape photometric} aperture is to achieve the optimal balance between flux loss and noises based on a relationship derived from the CCD equation (see \citealt{mer95}).  Figure \ref{fig:Fig2} shows how the optimum apertures vary with the stellar magnitude.  There is an obvious trend of decreasing aperture sizes with increasing magnitudes down to the faint magnitude limit in the example frame.

Once we measure the flux of each object with the optimum aperture, we need to apply the aperture correction for small apertures.  The aperture correction terms are estimated from the growth curve analysis of selected isolated bright stars (i.e., reference stars).  The average curve-of-growth for each frame is calculated by measuring the difference in magnitude between different pairs of apertures (up to 10 pixels aperture radius) and then an automatic correction is applied to all objects for each photometric aperture.  The use of a common aperture correction for each CCD assumes that there is no variation in the correction across the CCD.  This flux correction method gives nearly the same brightness within the measurement uncertainties for all apertures.  Any PSF variation across the CCD causes systematic errors, however, and we deal with this in Section 3.4 and Section 4.

\begin{figure}[t]
\centering
  \includegraphics[width=\linewidth, angle=0]{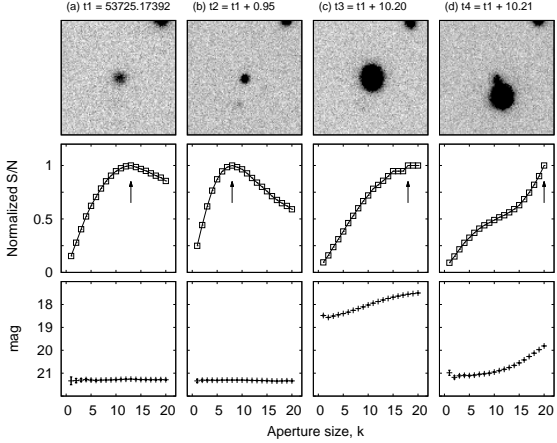}
  \caption{Example of multi-aperture photometry for one star (ID=10213 in the master source catalog) through epochs t1 to t4.  \emph{Top panels}: 100$\times$100 thumbnail images of the target star.  \emph{Middle panels}: normalized S/N as a function of aperture size.  \emph{Bottom panels}: aperture corrected magnitude as a function of aperture size.  The arrows represent the peak locations in the aperture-S/N diagram (see text for details).} 
  \label{fig:Fig3}
\end{figure}

We performed the multi-aperture photometry based on the concentric aperture photometry algorithm in DAOPHOT package \citep{ste87}, using several circular apertures (up to 10 pixels aperture radius) with a fixed sky annulus from 35 to 45 pixels.  The initial results of multi-aperture photometry are stored in ascii-format photometry tables, including the date of the observations (MJD), the pixel $(x, y)$ coordinates, the aperture-corrected magnitudes with errors for each aperture, the sky values and its errors.  Figure \ref{fig:Fig3} shows the details of the multi-aperture photometry for one star at different epochs.  In the former two epochs, the photometric apertures can be properly selected by the S/N cuts, while in the latter two epochs, S/N increases for lager apertures.  This unusual behavior is due to contamination by a moving object.  We automatically identifies similar unusual cases by the method of aperture indexing.

\begin{figure}[t]
\centering
  \includegraphics[width=\linewidth, angle=0]{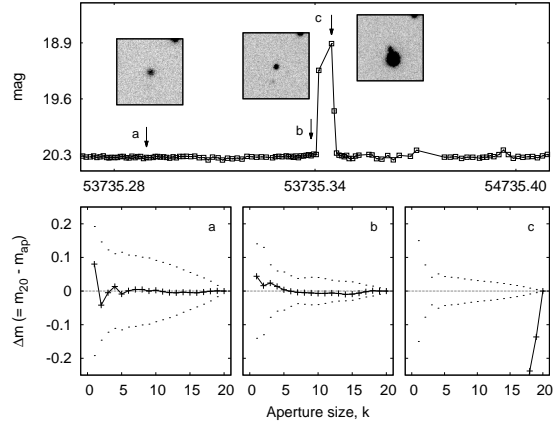}
  \caption{Typical example of multi-aperture indexing photometry.  Top panel is $r$-filter light curve of same star (ID=10213) as shown in Figure \ref{fig:Fig3}.  Bottom panels show the multi-aperture indexing scheme.  The $x$-axis is the aperture size and $the$ $y$-axis is the differential magnitude between pairs of apertures $\Delta m$(=$\acute{m_\mathrm{i,j,ref}}-\acute{m_\mathrm{i,j,k}}$).  We can see whether and at what aperture the differential magnitude (solid lines) begins to deviate from the model curve (dashed lines) for each epoch.}
  \label{fig:Fig4}
\end{figure}

\begin{figure*}[t]
\begin{center}
  \includegraphics[width=0.78\textwidth, angle=0]{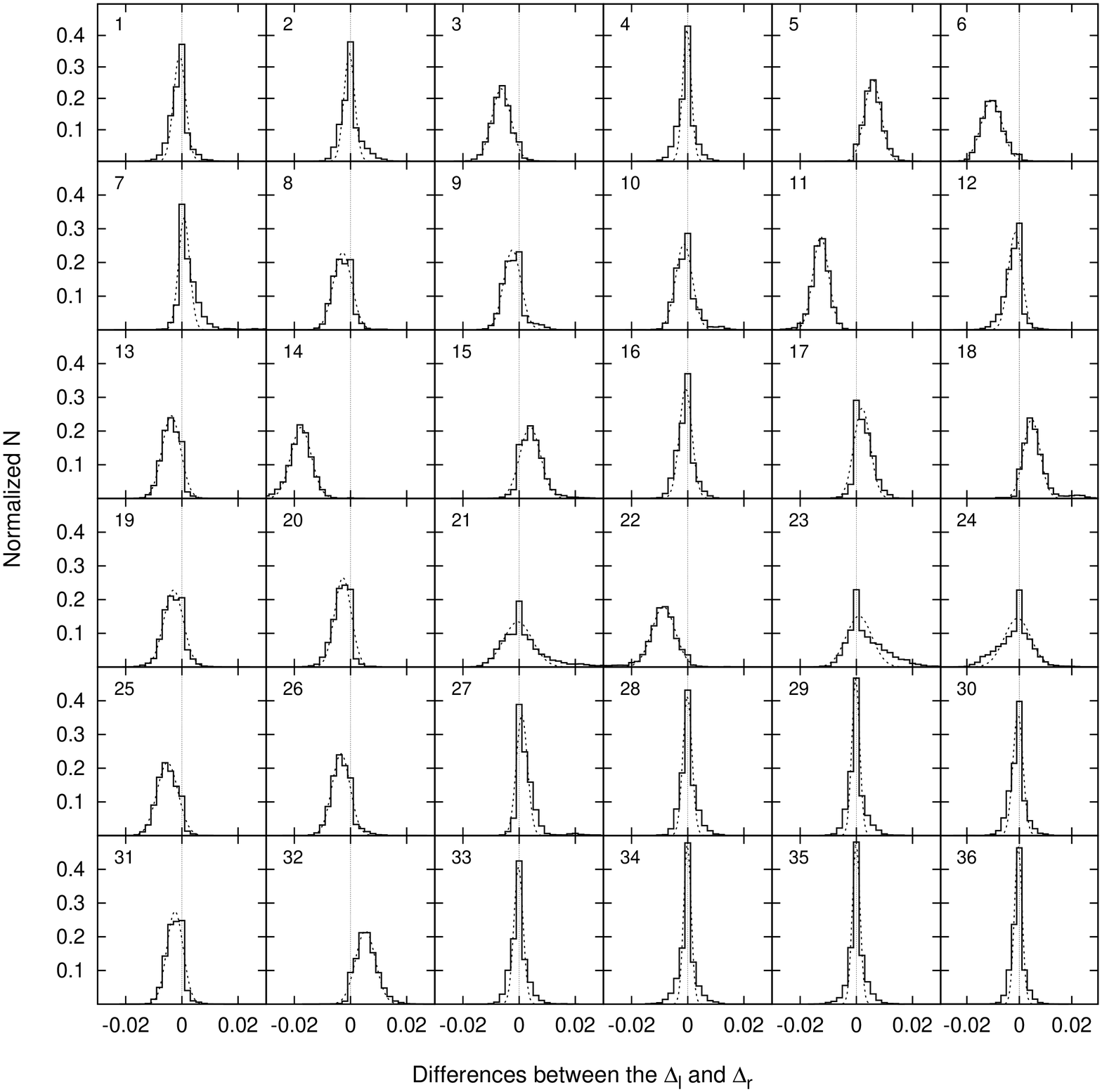}  
  \caption{Histogram of differences in the magnitude offsets between the left- and right-side of each chip ($\bar{\Delta}_\mathrm{l} - \bar{\Delta}_\mathrm{r}$) for the whole data set.  Each histogram is    normalized by the total number of data frames $N$ and is described by Gaussian distribution with different means and variances (dashed lines).}
    \label{fig:Fig5}
  \end{center}
\end{figure*}

\subsubsection{Determination of the Best Aperture with Indexing Method}
Our multi-aperture indexing method is similar to the basic concept of the discrete curve-of-growth method \citep{ste90}.  Each object is indexed based on the difference in aperture-corrected magnitude between pairs of apertures $\Delta m$(=$\acute{m_\mathrm{i,j,ref}}-\acute{m_\mathrm{i,j,k}}$) with mean trend for stars of similar brightness (see solid and dotted lines in bottom panels of Figure \ref{fig:Fig4}, respectively).  The aperture with a 10 pixel radius is used as the fixed reference aperture.  The mean trend is determined by computing the rms curve of the aperture correction values for all measured apertures, and used to evaluate whether magnitude at a given aperture significantly differs from the mean trend.  Since the rms value depends on the chosen magnitude interval, all stars are divided into groups according to their brightness in the individual frames.  We determine the rms curve for each magnitude group using an iterative $\sigma$-clipping until convergence is reached.  Objects lying within $\pm$3-$\sigma$ of the model curve are indexed as contamination-free, and those above $\pm$3-$\sigma$ as a contaminated source.  Figure \ref{fig:Fig4} shows that multi-aperture indexing guides us to throw out some photometric measurements if they are discrepant from the mean trend.  This approach also gives us a chance to recover a measurement that would be otherwise thrown out.  The problematic aperture can be simply replaced by one of the smaller apertures if it is indexed as contamination-free.  This help us make a full use of the information offered by the data.  

\subsection{Improved Photometric Calibration}
We present a new photometric calibration to convert the instrumental magnitudes onto the standard system, including a relative flux correction of the left and right half-region of each CCD chip.  As mentioned in the Section 3.1, MMT/Megacam shows the temporal variations in the gain between two amplifiers on each CCD, as well as between CCDs that are part of the same mosaic.  It may have been caused by unstable bias voltage of the CCD output drain which has a profound impact on the gain of the output amplifier.  The level of readout noise is also unstable between two amplifiers.  To correct for this effect, the photometric calibration needs to be performed individually for each amplifier region.

We use a sufficient number of (pre-selected) bright isolated stars as standard stars and compute the relative flux correction terms.  These terms were derived for each frame using the mean  magnitude offset ($\bar{\Delta}_\mathrm{l, r}$) of standard stars ($N_\mathrm{l, r}$) with respect to corresponding magnitudes ($m_\mathrm{i, j}$) in the master frame chosen as an internal photometric reference
\begin{mathletters}
\begin{equation}
\bar{\Delta}_\mathrm{l, r} = \frac{1}{N_\mathrm{l, r}}\sum m_\mathrm{i, j} - \acute{m_\mathrm{i, j}} + ZP_\mathrm{l, r},
\end{equation} where $\acute{m_{i, j}}$ is the aperture-corrected instrumental magnitudes in other frames and $ZP_\mathrm{l, r}$ is the photometric zero-points for the left- and right-side of each chip, respectively.  To calculate the zero-points, we solve a linear calibration relation of the form:
\begin{equation}
r - m_\mathrm{i, j}  = ZP_\mathrm{l, r} - 0.07 X + 0.107 (r - i),
\end{equation} where $r$ and $i$ are standard magnitudes from the photometric catalog of M37 \citep{har08a} and $X$ is an airmass term.
\end{mathletters} 
The fit is performed iteratively using a sigma-clipping method.  

Figure \ref{fig:Fig5} shows the difference in the magnitude offsets between the left- and right-side of each chip ($\bar{\Delta}_\mathrm{l} - \bar{\Delta}_\mathrm{r}$) for the whole data set, which is within $\pm0.02$ magnitude level for all 36 CCD chips.  The histograms are normalized by the total number of data frames $N = 4,730$ and are described by a Gaussian function with slightly different mean values and shapes (dashed line).  We clearly see a significant variation in difference between a pair of magnitude offsets for all CCDs.

\subsection{Field Distortion Correction}
\begin{figure}[t]
\centering
    \includegraphics[width=\linewidth,angle=0]{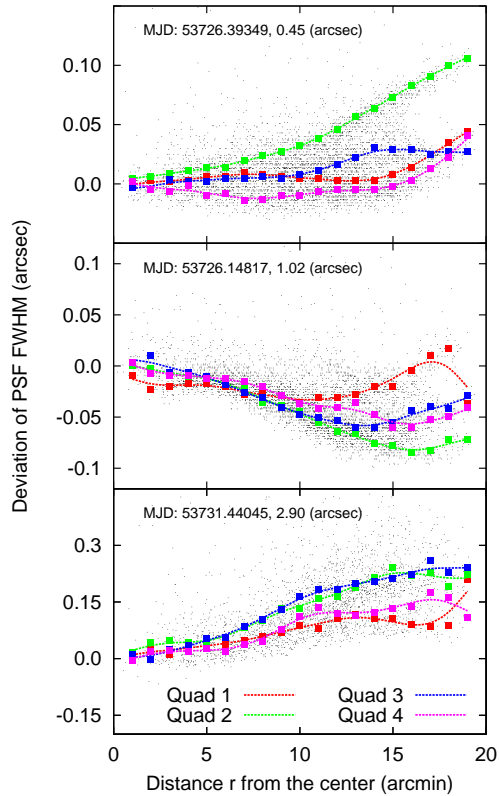}
    \caption{PSF FWHM variations as a function of distance from the image center for various seeing conditions.  For each quadrant (denoted by different colors), the dashed lines represent the weighted spline approximation of the median value of each distance bin (one arcmin).}
     \label{fig:Fig6}
\end{figure}

The photometric calibration for wide-field imaging systems is also affected by position-dependent systematic errors due to a PSF variation across the FOV (e.g., \citealp{ive07,hod09}).  We derive PSF variations across the FOV with the SExtractor package.  The change of the PSF shapes in the image plane is represented by spatial distribution of PSF FWHM values for several bright stars, with parameters of \texttt{CLASS$\_$STAR} $>$ 0.9, \texttt{MAGERR$\_$AUTO} $<$ 0.01 mag, and \texttt{FLAGS} = 0 (i.e.,  isolated point sources with no contamination).  Note that the PSF FWHM values are defined as the diameter of the disk that contains half of the object flux based on a circular Gaussian kernel.  Figure \ref{fig:Fig6} presents the variation of the PSF FWHM as a function of distance from the image center for various seeing conditions.  For each quadrant, the dashed lines represent the weighted spline approximation of the median value of each distance bin (1 arcmin).  The result shows that the PSF FWHM varies significantly as a function of position on the single-epoch image frames and variations are at the level of $\sim$10\% to 20\% ($0\arcsec.1-0\arcsec.2$) across the FOV.  As the field distortion is not negligible from the center of field to its edges, such variations limit the accuracy of stellar photometry.

To address this issue, we perform a 2D polynomial fitting technique.  For each frame, the correction terms are described by a linear or quadratic polynomial depending on the position $(x,y)$ only.
\begin{widetext}
\begin{equation}
\begin{split}
\Delta \vec{m}_\mathrm{c,k}(x,y,m_\mathrm{err})_\mathrm{lin} & = c_\mathrm{0k} + c_\mathrm{1k}\vec{x} + c_\mathrm{2k}\vec{y},\\
\Delta \vec{m}_\mathrm{c,k}(x,y,m_\mathrm{err})_\mathrm{qud} & = c_\mathrm{0k} + c_\mathrm{1k}\vec{x} + c_\mathrm{2k}\vec{y} + c_\mathrm{3k}\vec{x}^{2} + c_\mathrm{4k}\vec{y}^{2} + c_\mathrm{5k}\vec{xy},
\end{split}
\end{equation} 
\end{widetext}where $x, y$ are the pixel coordinates of $N$ bright isolated stars, $m_\mathrm{err}$ is the statistical weight in the fitting procedure, $\vec{c}_\mathrm{k}$ are the sets of polynomial coefficients for each aperture size, and $\Delta \vec{m}_\mathrm{c, k}(x,y,m_\mathrm{err})$ are the difference in magnitude between the reference aperture and $k$ aperture, $\Delta \vec{m}_\mathrm{c, k}(x,y) = \vec{m}_\mathrm{c, 20}(x,y) - \vec{m}_\mathrm{c, k}(x,y)$, at the position $(x,y)$ for each chip $c$.  We derived the optimal parameter values from a nonlinear least-squares fit using the Levenberg--Marquardt algorithm and automatic differentiation,\footnote{We used a \texttt{LeastSquareFit} module provided in the scientific python package (\url{http://www.scipy.org/}).} and choose between two models that best fit the data.

\begin{figure*}[t] 
   \centering
   \subfloat[]{\includegraphics[width=0.34\linewidth, angle=-90]{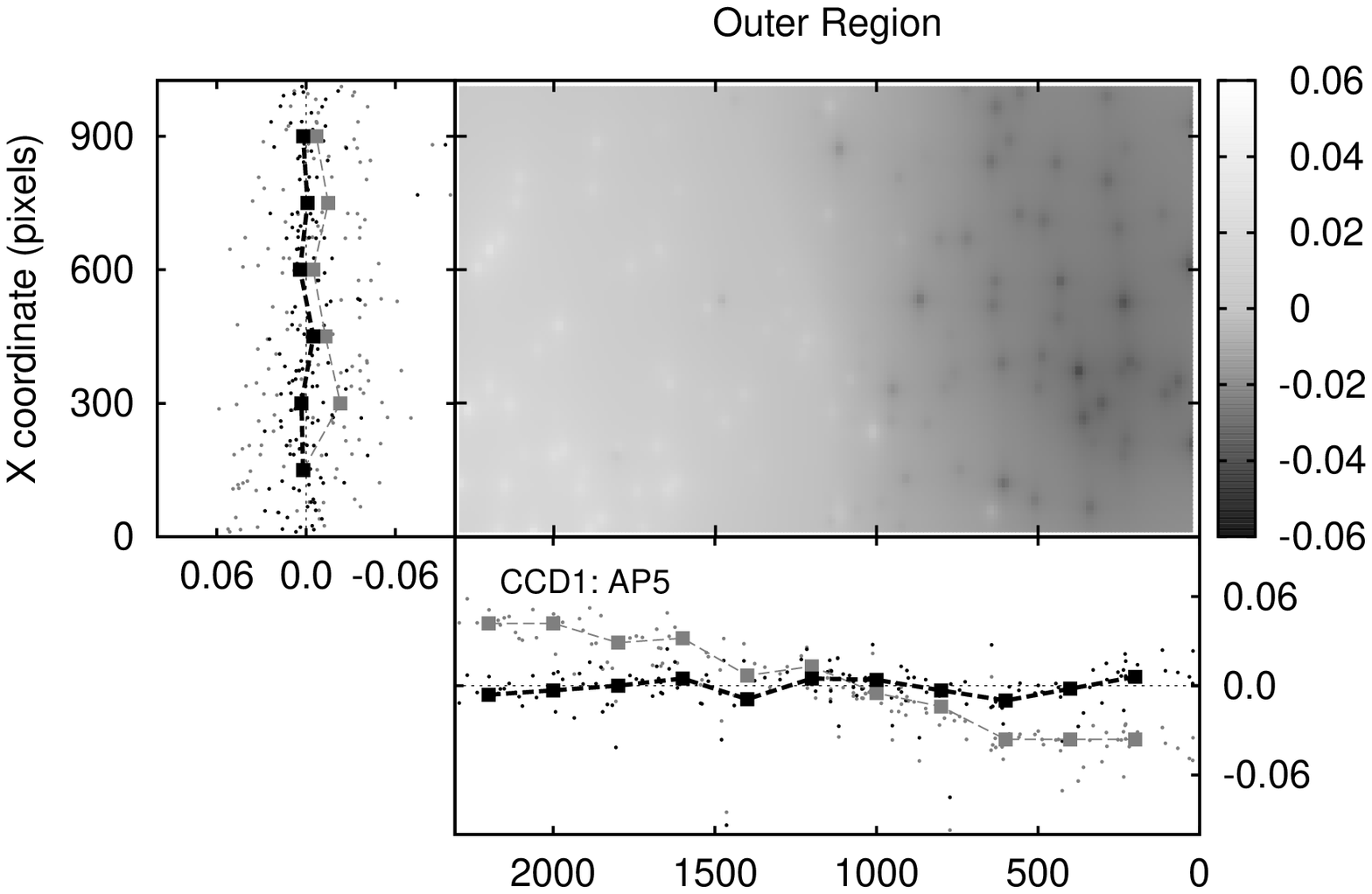}}  
   \subfloat[]{\includegraphics[width=0.34\linewidth, angle=-90]{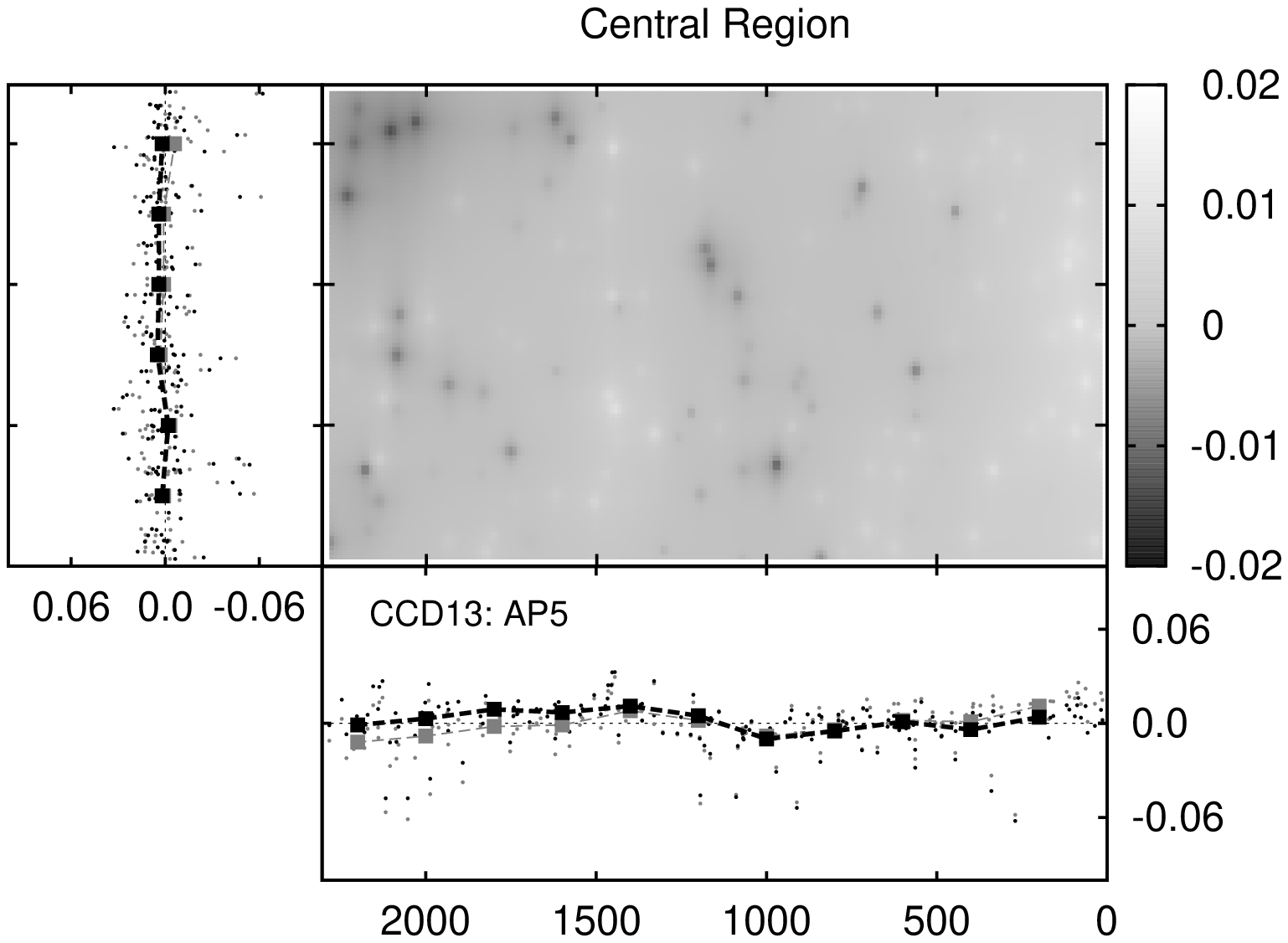}}\vspace*{-1.5em}
   \subfloat[]{\includegraphics[width=0.34\linewidth, angle=-90]{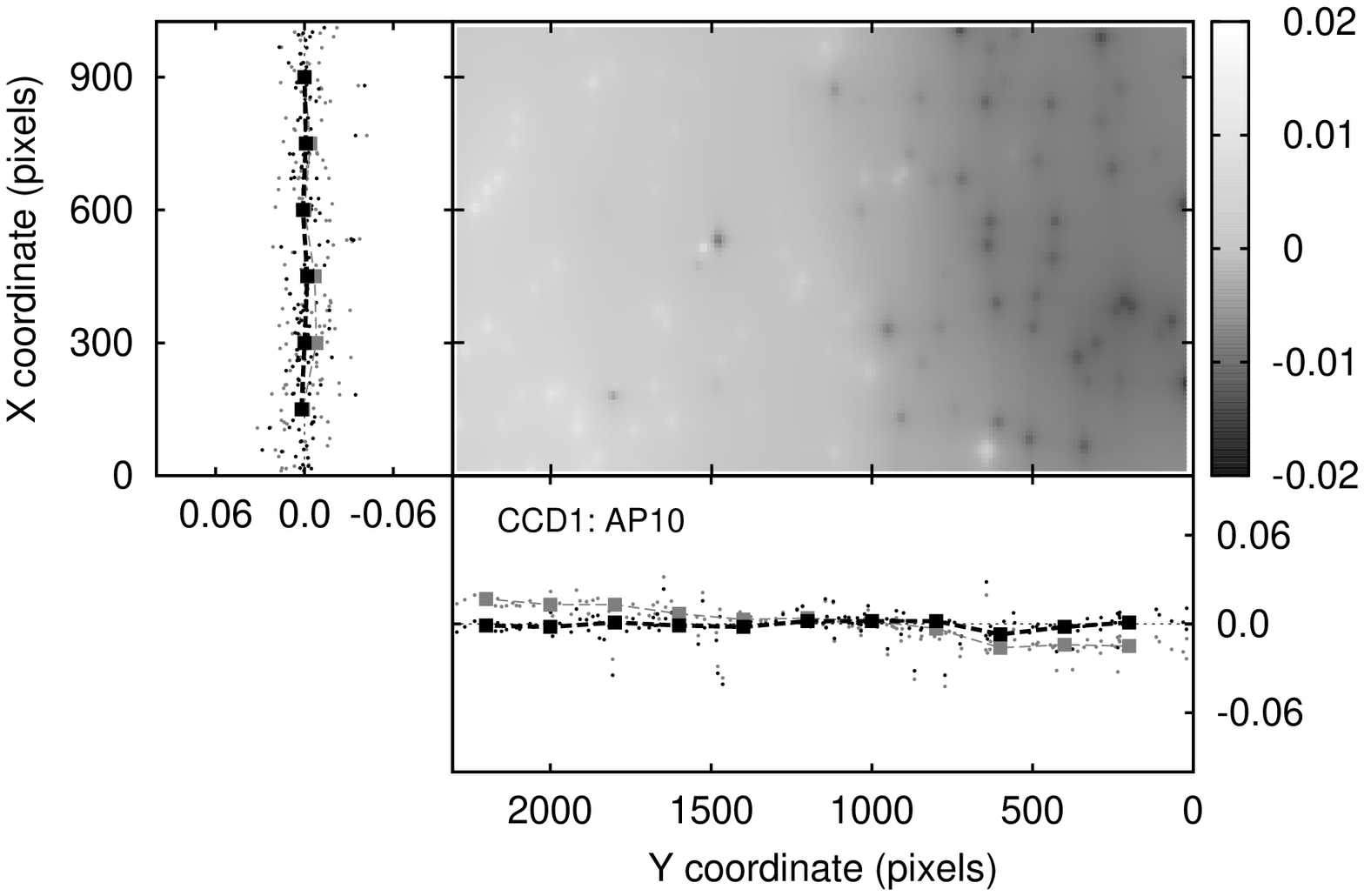}}  
   \subfloat[]{\includegraphics[width=0.34\linewidth, angle=-90]{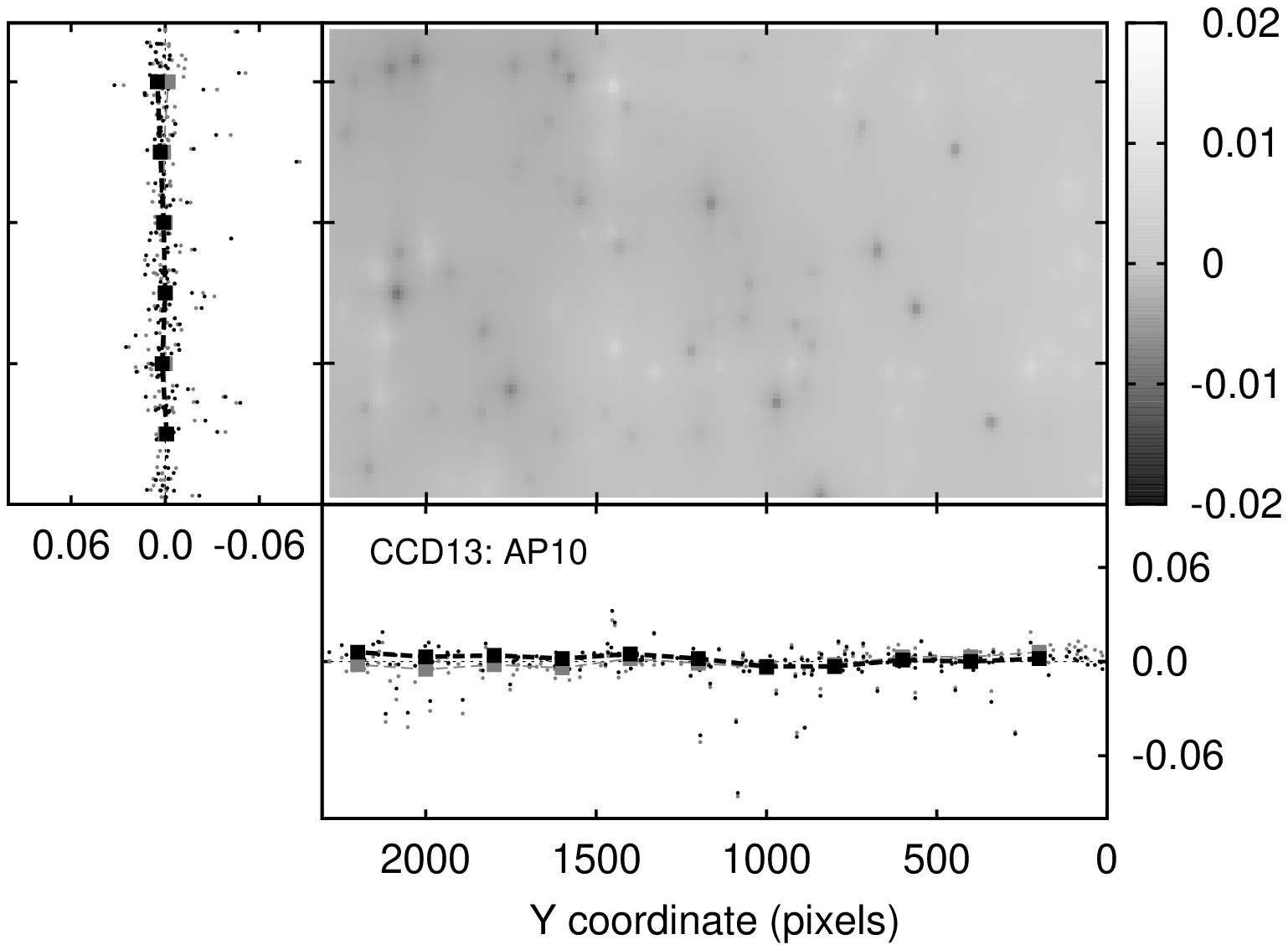}}\vspace*{-1.3em}
   \caption{Example of the position-dependent magnitude offsets ($gray$ symbols) and the distortion correction by 2D polynomial fitting method ($black$ symbols) for selected CCD chips.  For the outer ($left$) and the central ($right$) region of the mosaic, we compare the magnitude offsets between the reference aperture and the relatively smaller apertures (e.g., AP5, AP10) as a function of $(x,y)$ coordinates.  The variation is usually more significant in the $y$-direction than in the $x$-direction, especially for the case of aperture photometry performed with small apertures.}
   \label{fig:Fig7}
\end{figure*}

\begin{figure}[t]
\centering
  \includegraphics[width=\linewidth, angle=0]{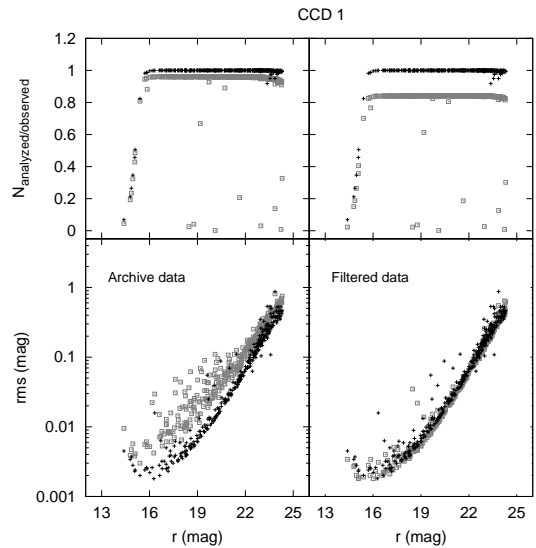}  
  \caption{Comparison of data recovery rate (\emph{top} panels) and rms photometric precision of light curves (\emph{bottom} panels) for the selected stars in the outer part (CCD 1) of the field of view.  The black points are from the re-calibrated light curves, while the gray points are from the raw (left) and filtered (right) light curves in archive, respectively.  Our results use nearly 100\% of observed data and reach comparable accuracy without any filtering.}
  \label{fig:Fig8}
\end{figure}

Figure \ref{fig:Fig7} shows the field-dependent magnitude offsets and the distortion correction by 2D polynomial fitting method for one example mosaic CCD.  For the outer and the central region of the mosaic, we compare the magnitude offsets between the reference aperture and the relatively smaller apertures as a function of $(x,y)$ coordinates.  Here $x$-axis is in the declination direction and $y$-axis is opposite to the right ascension direction.  We find that the magnitude difference depends on position $(x, y)$ and is most discrepant in the outer part of the FOV.  This effect is usually more significant in the $y$-direction than in the $x$-direction, especially for the case of aperture photometry performed with small apertures.  The correction for field-dependent PSF variation reduces the initial $\sim$10\% variation (gray lines) to less than $\sim$1\% (black lines).

\begin{figure}[t]
\centering
  \includegraphics[width=\linewidth, angle=0]{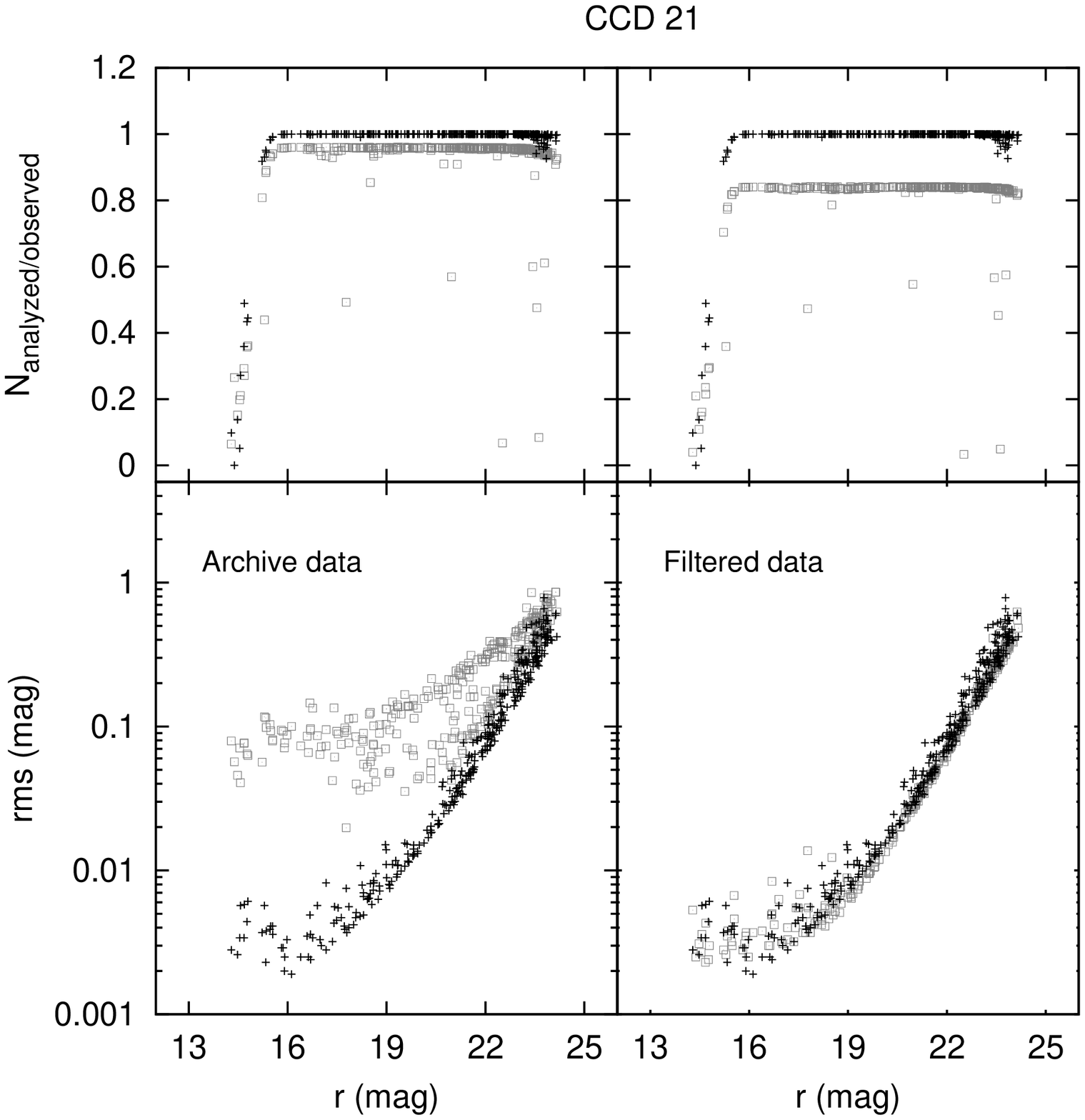}  
  \caption{Comparison of data recovery rate (\emph{top} panels) and rms photometric precision of light curves (\emph{bottom} panels) for the selected stars in the central part (CCD 21) of the field of view.  The plotted symbols and notations are the same as Figure \ref{fig:Fig8}.}
  \label{fig:Fig9}
\end{figure}

\subsection{Photometric Performance Diagnostics}
\subsubsection{Comparison with Archival Data}
We compare the photometric performance of the re-calibrated light curves with the non-de-trended archival light curves\footnote{Note that the archival data have not been trend-filtered, but \citet{har09} used the de-trended archival light curves as part of the transit detection process in their paper IV of the prior analysis.} by means of the two representative measures: (i) the rms photometric precision $\sigma_\mathrm{ph}$, and (ii) the data recovery rate $N_\mathrm{recovery}$.  The former is defined as the standard deviation of light curves around the mean value as a function of $r$ magnitude:
\begin{equation}
\sigma_\mathrm{ph}(N_\mathrm{j},\bar{m}_\mathrm{j})  = \sqrt{\sum^{N_\mathrm{j}}_\mathrm{p=1}\frac{(m_\mathrm{p}-\bar{m}_\mathrm{j})^{2}}{N_\mathrm{j} - 1}},
\end{equation} where $N_\mathrm{j}$ is the number of data points in each light curve, $m_\mathrm{p}$ is the observed magnitude, and $\bar{m}_\mathrm{j}$ is the mean magnitude of the object $j$, and the latter refers to the number of \emph{analyzed} data frames normalized by the total number of observed data frames $N$ for each object.  In typical cases, the data recovery rate should be near unity in the bright magnitude regime and decreases with magnitude for fainter objects.  For comparison, we decided to select light curve samples which show either no significant variability or seeing-correlated variations induced by image blending.  We remove all known variable stars from the sample list based on a new catalog of variable stars in M37 field (S.-W. Chang et al. 2015, hereafter Paper II).  To remove the light curves of blended objects, we use an empirical statistical technique to quantify the level of blending by looking for seeing-correlated shifts of the object from its median magnitude \citep{irw07}.
\begin{displaymath}
b = \frac{\chi^{2}-\chi_\mathrm{poly}^{2}}{\chi^{2}},
\end{displaymath}
where
\begin{equation}
\chi^{2} = \sum^{}_\mathrm{p}\frac{(\sum m_\mathrm{p}-\bar{m}_\mathrm{j})^{2}}{\sigma_\mathrm{p}^{2}}
\end{equation} for light curve points $m_\mathrm{p}$ with uncertainties $\sigma_\mathrm{p}$, and $\chi_\mathrm{poly}^{2}$ is the same statistic measured with respect to a fourth order polynomial in FWHM fitted to the data.  We adopt the value $b < 0.4$ for the selecting light curves with no blending.  The last selection criterion is that the light curves must exit both in the archive and our database.

In the bottom panels of Figure \ref{fig:Fig8} and Figure \ref{fig:Fig9}, we plot the rms photometric precision of light curves for the two Megacam CCD chips in the outer (CCD 1) and central (CCD 21) part of the FOV, respectively.  The black points show the rms values of the re-calibrated light curves, while the gray points are for the raw and filtered light curves in archive.  The first impression from this comparison is that the typical rms scatter is overestimated from the raw light curves because of many outliers in the photometric data (bottom left panels).  For the better results, these light curves were filtered out in two steps: (i) clipping 5-$\sigma$ outliers from each light curve and (ii) removing every data points that are outliers in a large number of light curves \citep{har08b}.  In the second step, the outlier candidates are estimated by choosing a cutoff value for each CCD chip.  The cutoff value is defined as the fraction of light curves for which a given image is a 3-$\sigma$ outlier.  This filtering was applied to remove bad measurements due to image artifacts or poor conditions, which were previously thought to be unrecoverable, but resulting data loss is up to 20\% of the total number of data points (bottom right panels).  As shown in the top panels of the two figures, the data recovery rate for the re-calibrated light curves is close to 100$\%$ over a wide range of magnitude and it appears to be more complete compared with the raw (top left panels) and filtered (top right panels) light curves.  At bright magnitudes ($r < 16$) the data recovery rate does not reach 100$\%$ because the exposure time was chosen to be saturated at a magnitude of $r\sim15$.  This comparison proves that our approach is a powerful strategy for improving overall photometric accuracy without the need to throw out many outlier data points.

\begin{figure*}[t]
   \centering
  \subfloat[]{\includegraphics[width=0.35\linewidth, angle=-90]{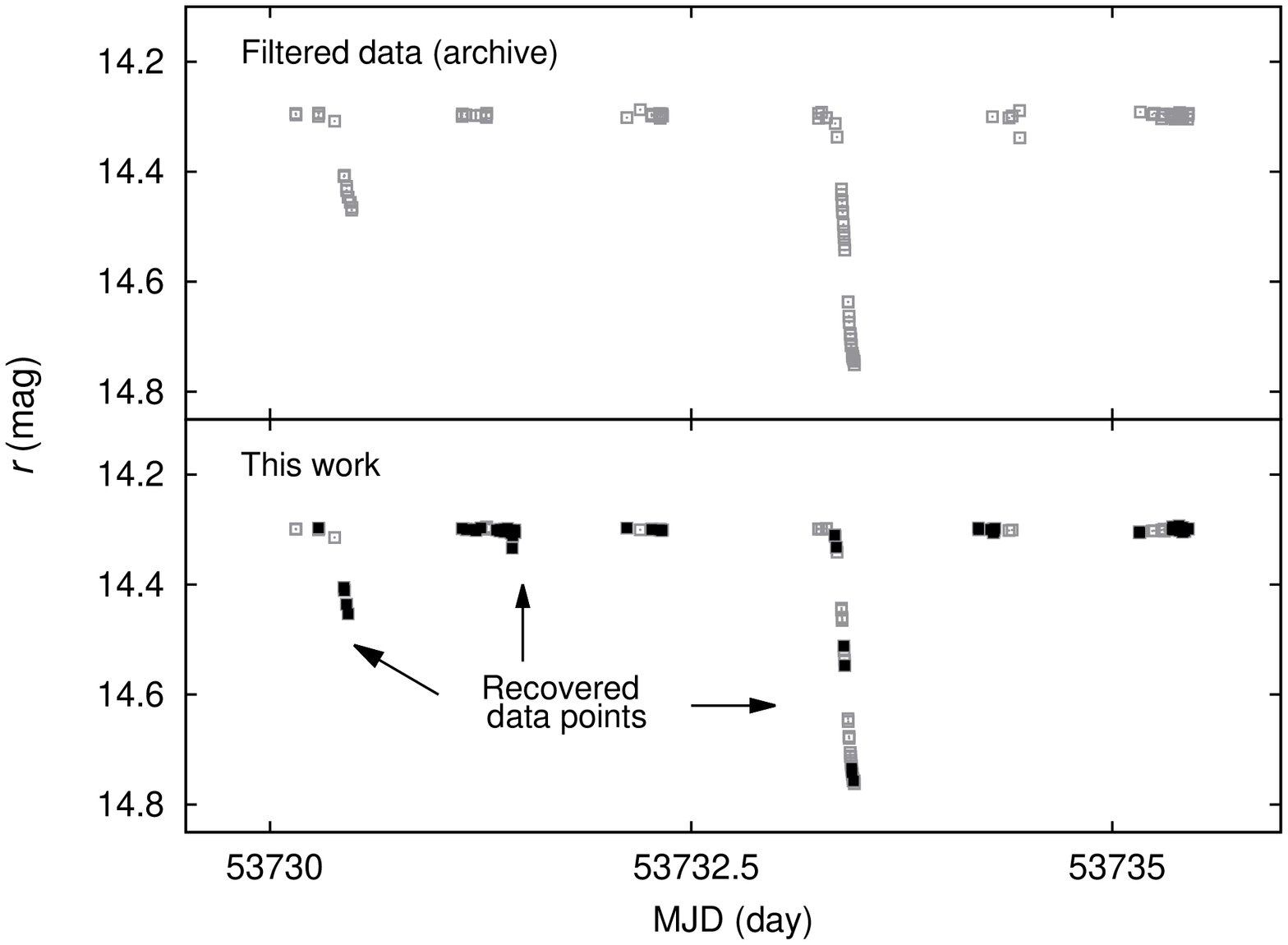}}  
  \subfloat[]{\includegraphics[width=0.35\linewidth, angle=-90]{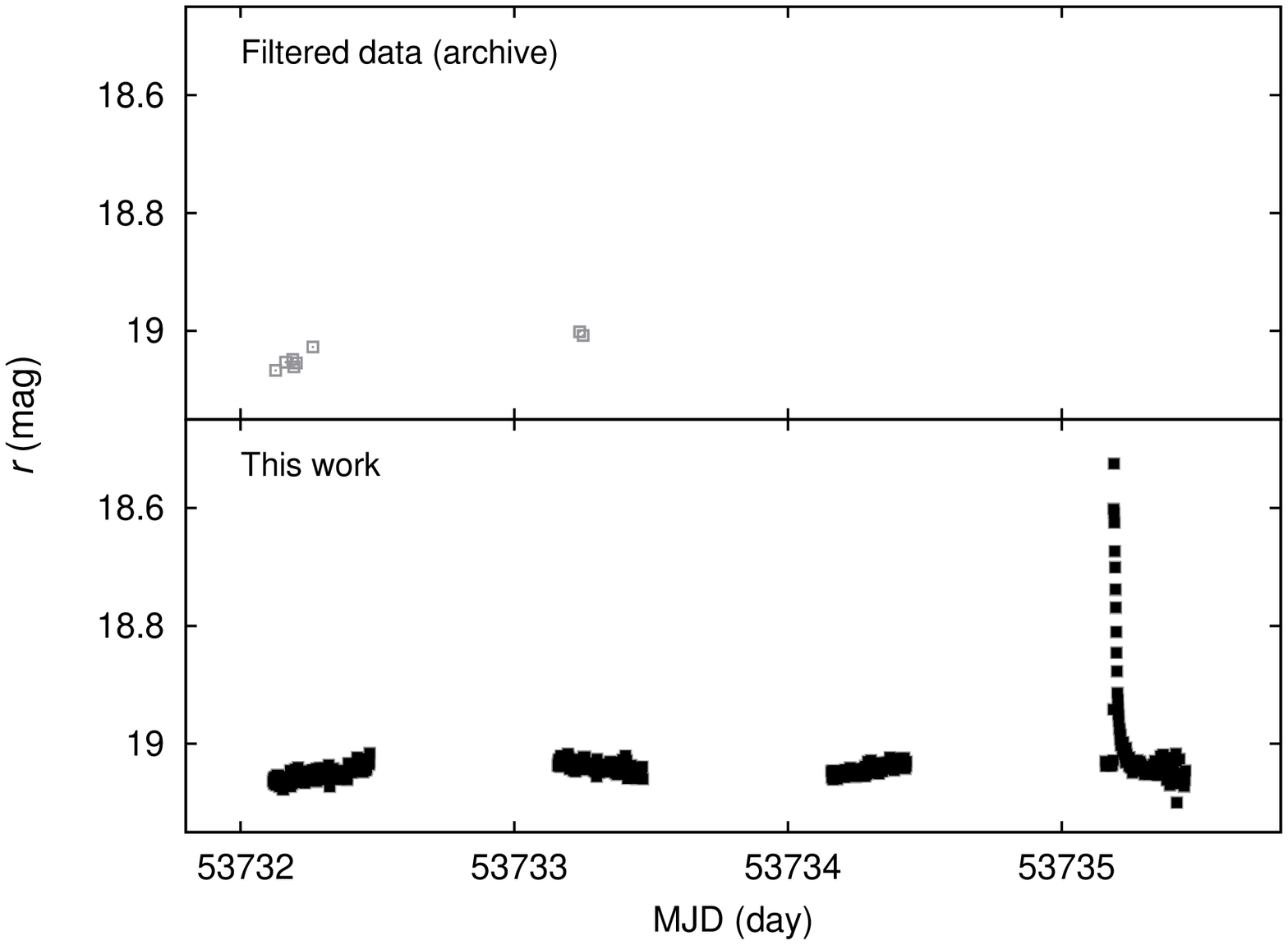}}\vspace*{-1.3em}
  \subfloat[]{\includegraphics[width=0.35\linewidth, angle=-90]{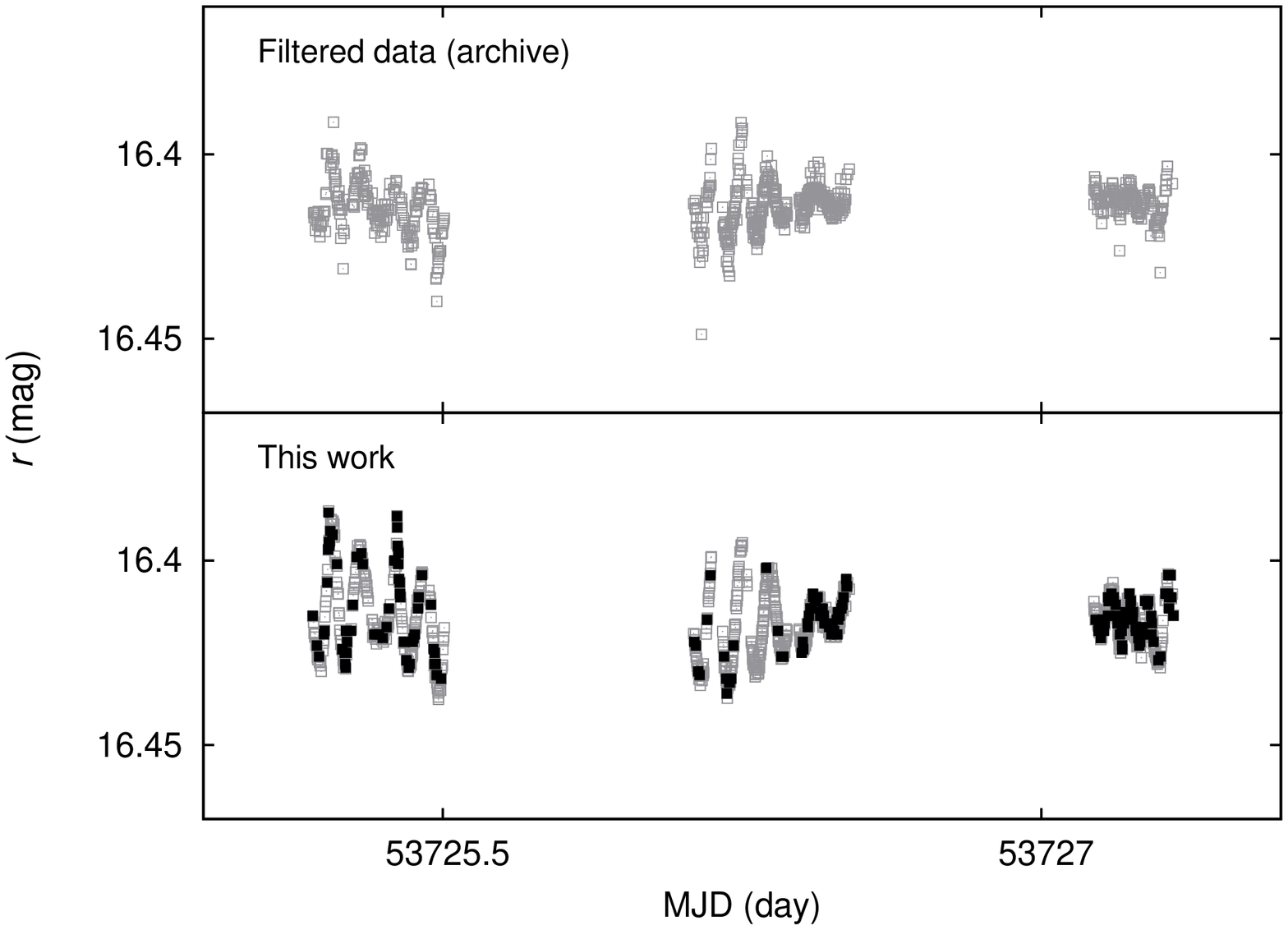}} 
  \subfloat[]{\includegraphics[width=0.35\linewidth, angle=-90]{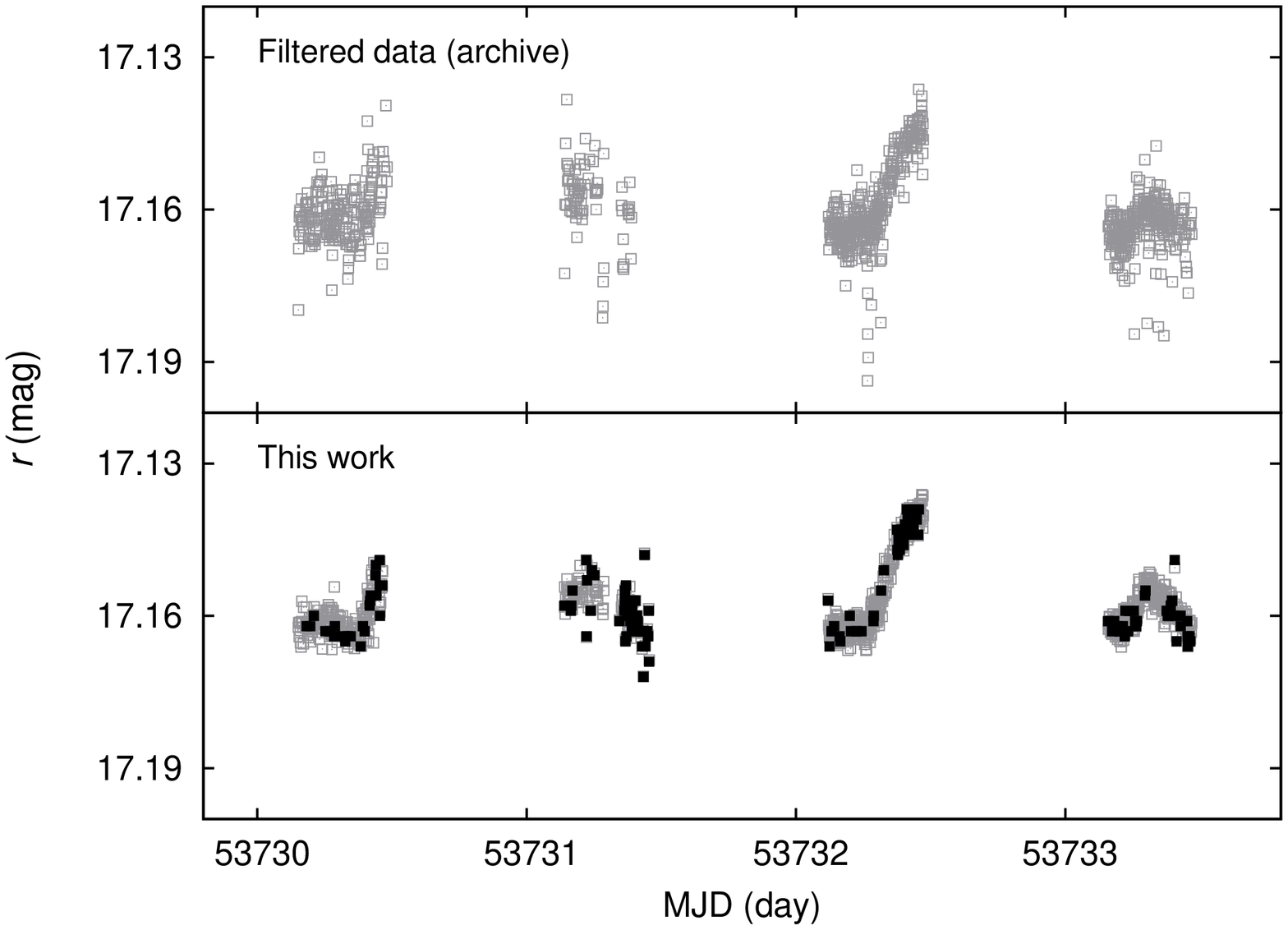}}\vspace*{-1.3em}
   \caption{Light curves for selected variable stars from Hartman et al. (2008b, top panels) and from this work (bottom panels), respectively.  The recovered data points are marked with black squares.  From upper left to lower right: ID=310010 (detached eclipsing binary star), ID=310139 (rotating variable star with flare), ID=100031 (pulsating variable star), and ID=100046 (aperiodic variable star).}
   \label{fig:Fig10}
\end{figure*}

Finally, we compare the light curves themselves for selected variable stars between the archive and our own.  This comparison serves to illustrate how the photometric precision and data recovery rate of the time-series data affect the ability to address a variety of variability characteristics.  Figure \ref{fig:Fig10} shows a direct comparison with the filtered light curves (top panels) and our re-calibrated light curves (bottom panels) for four variable stars.  It is shown that our method recovers more data points (black) from the same data set of images. 

\begin{figure}[t]
\centering
   \includegraphics[width=\linewidth, angle=0]{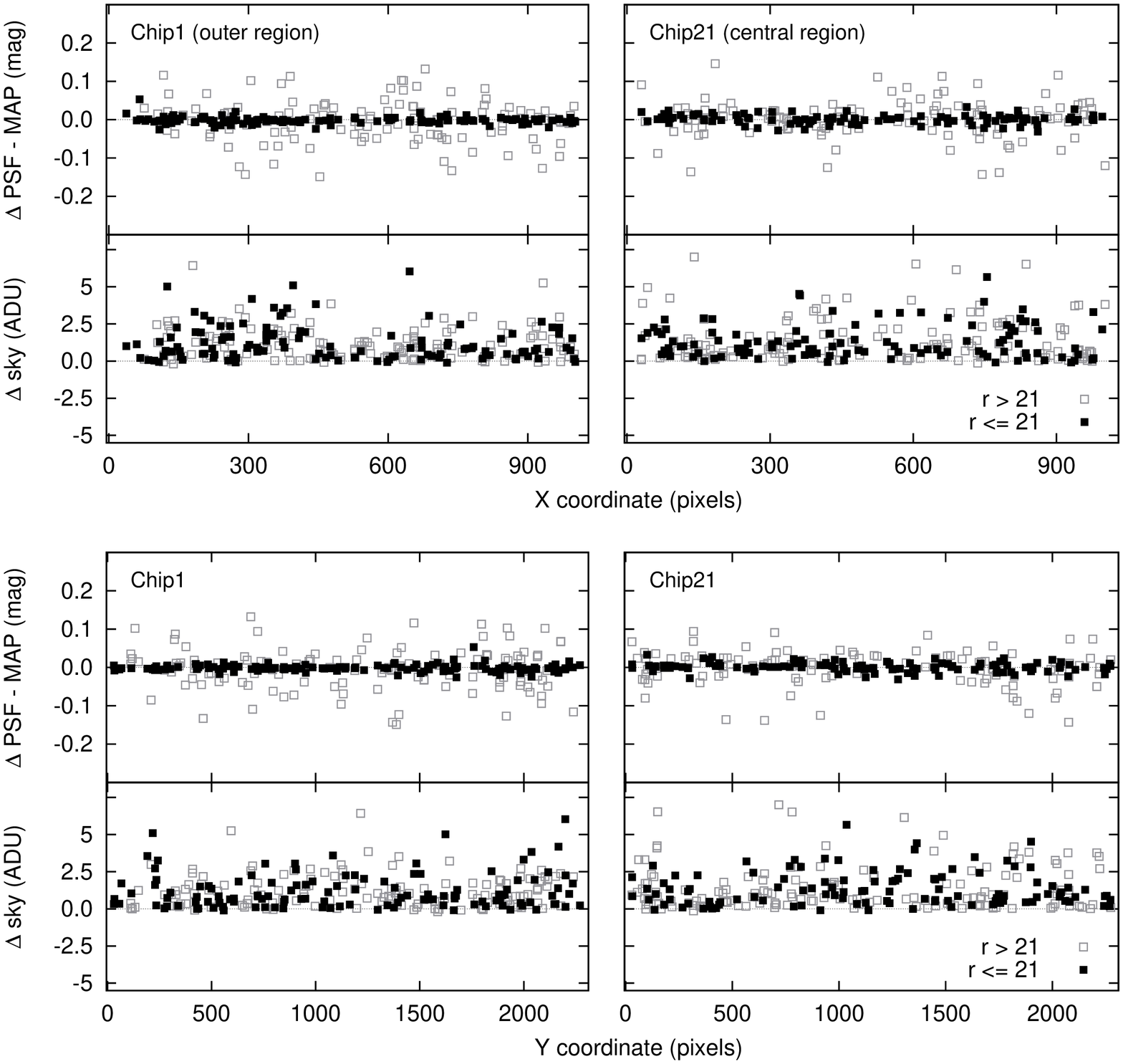}
    \caption{Example of residual magnitudes and sky values between our multi-aperture photometry and the PSF-fitting photometry as a function of $x$ and $y$ coordinates, respectively, for selected CCD chips.  The filled squares are relatively bright stars with $r \le 21$ mag, while the open squares are stars with $r > 21$ mag.}
   \label{fig:Fig11}
\end{figure}

\subsubsection{Comparison with PSF-fitting Photometry}
In order to further investigate possible systematics in our approach, we conducted PSF-fitting photometry with DAOPHOT II and ALLSTAR \citep{ste87}.  For each mosaic frame, we select bright, isolated, and unsaturated stars to make the PSF model varying quadratically with $(x,y)$ coordinates.  After PSF modeling, we run ALLSTAR to perform iterative PSF photometry of all detected sources in the frame with initial centroids set to the same values used for our own photometry.  We then calculated aperture corrections using the package DAOGROW \citep{ste90} after subtraction of all but PSF stars, which creates aperture growth curves for each frame and then integrates them out to infinity to obtain a total magnitude for each PSF star.   The final step is to convert the instrumental magnitudes into the standard photometric system.  For each frame, the initial zero-point correction is applied by correcting the magnitude offset with respect to the master frame. This places photometry for all frames on a common instrumental system.  Following the same procedure in Section 3.3, the photometric calibration is performed individually for each amplifier region. 

Figure \ref{fig:Fig11} shows the residual magnitudes and sky values between our multi-aperture photometry and the PSF-fitting photometry as a function of position in the selected CCD chips.  There are no position-dependent trends in the magnitude residuals.  For the brighter stars with $r \le 21$ mag, the rms magnitude difference between the two methods is very small ($\Delta_\mathrm{chip 1}$ = $-0.001\pm0.009$ and  $\Delta_\mathrm{chip 21}$ = $0.001\pm0.012$, respectively), while for the relatively faint stars the rms difference is somewhat larger ($\Delta_\mathrm{chip 1}$ = $-0.002\pm0.055$ and $\Delta_\mathrm{chip 21}$ = $0.006\pm0.050$, respectively).  The results of this example indicate that we can reliably correct for the PSF variations by our calibration procedures.  Meanwhile, our sky values are slightly higher than the ALLSTAR sky values, but not to the degree that can seriously affect photometric measurements.

\begin{figure}[t]
\centering
   \includegraphics[width=\linewidth, angle=0]{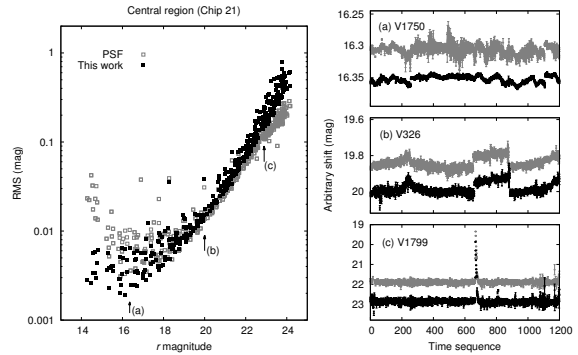}
    \caption{Comparison of the photometric precision (rms) for our multi-aperture photometry and for the PSF-fitting photometry as a function of the $r$ magnitude in the central region of the open cluster M37 (left panel).  The arrows indicate the three different magnitude levels from bright to faint in our sample shown in the right panels.  Note that the variable object IDs are taken from the new variable catalog of the M37 (see Paper II).}
   \label{fig:Fig12}
\end{figure}

Figure \ref{fig:Fig12} shows a comparison of the rms dispersion of the light curves obtained with our photometry with respect to the that of the PSF-fitting photometry in the central region of the open cluster M37.  We only compare the light curves of non-blended objects as described in Section 3.5. Our multi-aperture photometry does not reach the same level of precision as PSF-fitting photometry for the faintest stars, while the PSF-fitting approach results in poorer photometry for bright stars.  As shown in the right panels of Figure \ref{fig:Fig12}, it is clear that our photometry tend to have smaller measurement errors with respect to the PSF photometry for the bright stars.

\begin{figure}[t]
\centering
   \includegraphics[width=\linewidth, angle=0]{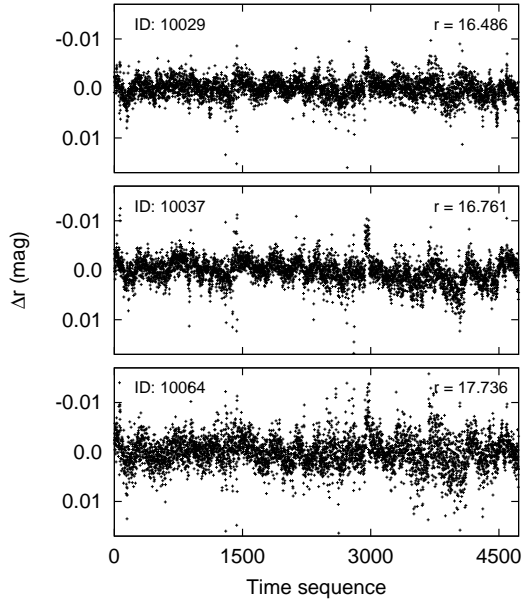}
    \caption{Example of shared systematic trend in the light curves of isolated bright stars.  The numbers on the upper left- and right-side of each panel show the star's identification number and the mean value of magnitude, respectively.  The numbers on the $x$-axis indicate the corresponding timestamps in each frame.  The deviations from the mean value, $\Delta r$, are less than $\pm$0.01 mag level (with a rms values of 0.0021, 0.0025, and 0.0033 mag from the top panel down).  These stars show a similar pattern of light variations over the observation span.}
   \label{fig:Fig13}
\end{figure}

\section{TEMPORAL TRENDS IN THE RE-CALIBRATED LIGHTCURVES}
From a visual inspection of the re-calibrated light curves in the same CCD chip, we found that some light curves tend to have the same pattern of variations over the observation span (Figure \ref{fig:Fig13}).  This kind of systematic variation (i.e., \emph{trend}) is often noticed in other studies.  For example, the importance of minimizing known (or unknown) systematics have been recognized by several exo-planet surveys because planet detection performance can be easily damaged by them (e.g.,\citealt{kov05,tam05,pon06}).  Also space-based time-series data (e.g., {\sc CoRoT} and {\sc Kepler}) are no exception to this behavior although it is completely free from systematics caused by the turbulent atmosphere.  Most of the raw light curves are affected by a secular (or a sudden) variation of flux without any obvious physical reason \citep{maz09,mis10,jen10}.

\begin{figure}[t]
\centering
  \subfloat[]{\includegraphics[width=0.46\textwidth,height=0.15\textheight, angle=0]{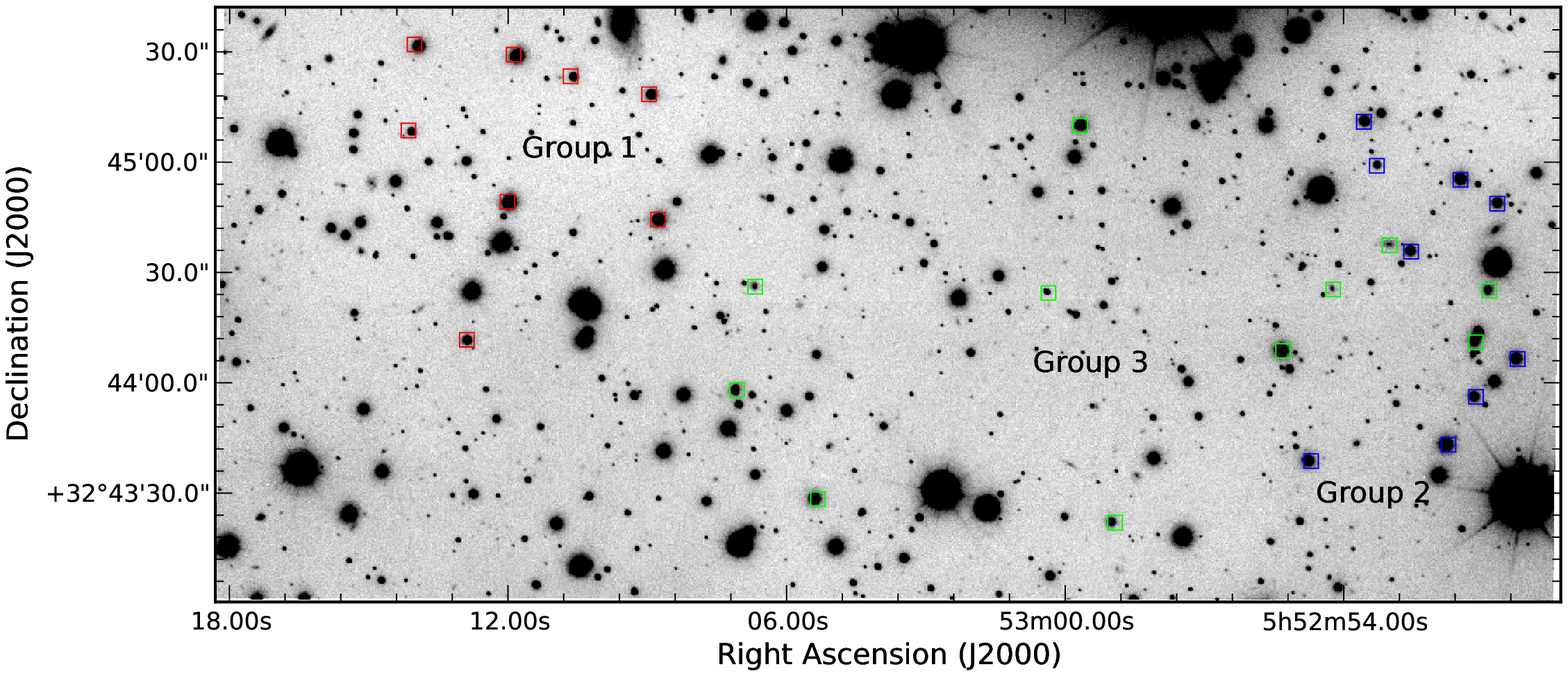}}\vspace*{-1.3em}
  \subfloat[]{\includegraphics[width=0.47\textwidth, height=0.23\textheight, angle=0]{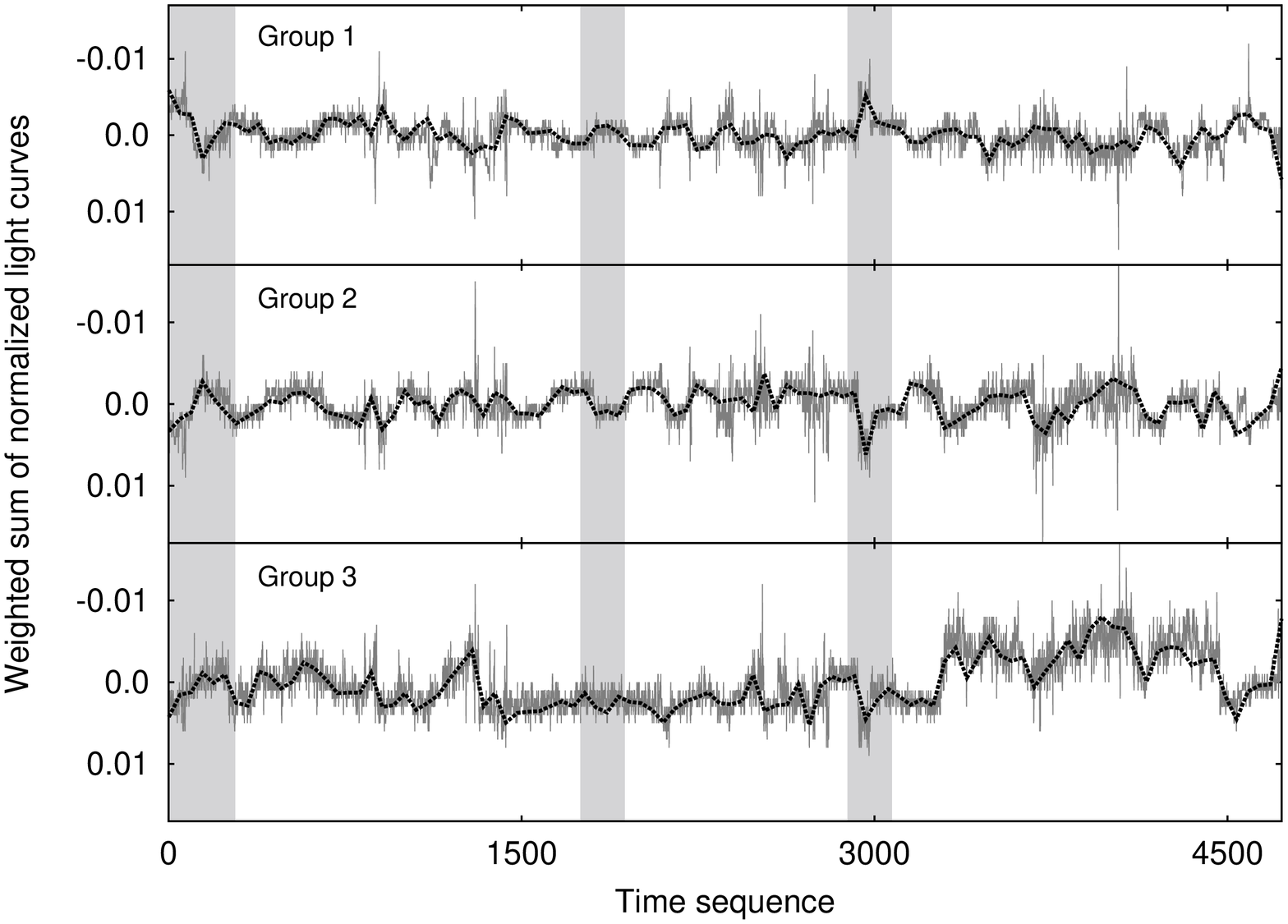}}\vspace*{-1.3em}
  \caption{\emph{Top panel}: spatial distribution of the most prominent trend groups on the CCD plane which covers the small areas of the sky ($6\arcmin.14\times2\arcmin.73$).  The colored dots represent the most systematic stars classified as group 1 (red), group 2 (blue), and group 3 (green), respectively.  One interesting property of trends is that they are localized within a CCD frame where stars are isolated from each other.  \emph{Bottom panel}: systematic features extracted from the selected light curves.  These are calculated by the weighted sum of normalized light curves from each group.}
    \label{fig:Fig14}
\end{figure}

In order to check the properties of temporal systematics, we examined the correlation coefficients as measure of similarity between two light curves $i$ and $j$ obtained from a single CCD chip (CCD 1).
\begin{equation}
C_\mathrm{ij}=\frac{1}{N-1}\frac{\sum^{n}_\mathrm{t=1} L_\mathrm{i}(t)L_\mathrm{j}(t) - n\overline{L}_\mathrm{i}\overline{L}_\mathrm{j}}{\sigma_\mathrm{i}\sigma_\mathrm{j}},
\end{equation} where $L(t)$ is the flux of each star at time $t$, $N$ is the total number of measurements, $\overline{L}$ is the mean flux of each star, and $\sigma$ is the standard deviation of $L(t)$.  This comparison is a point-by-point comparison and is done for every pair of light curves in the data set.  The resultant similarity matrix can be used to identify correlated pairs of light curves and to determine which light curve is least like all other light curves (e.g., \citealt{pro06}).  After that, we selected stars showing most systematics based on a hierarchical clustering method with the correlation coefficients (See \citealt{kim09}, for more details).  Figure \ref{fig:Fig14} represents spatial distribution of the most prominent trend groups on the CCD plane (top panel) and its strongly correlated features determined by the weighted sum of normalized light curves (bottom panel).  There are two interesting features in this figure: the first one is that each trend covers only a certain part of the sky area and the second one is that some portions of neighboring trends show different variation patterns even at the same moment in time (shaded gray region in the figure).  In particular, we found an anti-correlated variation for the trends between the group 1 ($g_{1}$) and the group 2 ($g_{2}$), so we might expect to find possible noise sources that are responsible for these discrepancies.  Why the trends are different and localized within a single CCD frame is a subject of further study, but it is probably related to subtle changes in point spread function and sky condition within the detector FOV.

\begin{figure}[t]
\centering
 \includegraphics[width=\linewidth,angle=0]{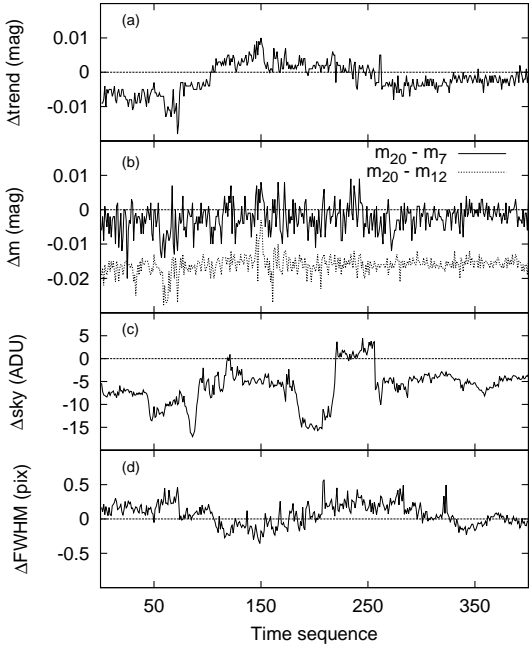}
 \caption{Example subset of data for comparing the trends with several parameters from the two groups ($g_\mathrm{1}$ and $g_\mathrm{2}$) of stars.   The numbers on the $x$-axis indicate the corresponding timestamps between the frame 1 and the frame 400.   From the top to bottom,  the panels show the average difference of trend, flux concentrations in which we applied a $-0.02$ mag shift to the $m_\mathrm{20}-m_\mathrm{12}$ value, sky level, and PSF FWHM value between the two groups.}
  \label{fig:Fig15}
\end{figure}

\begin{figure}[t]  
\centering
    \includegraphics[width=\linewidth, angle=0]{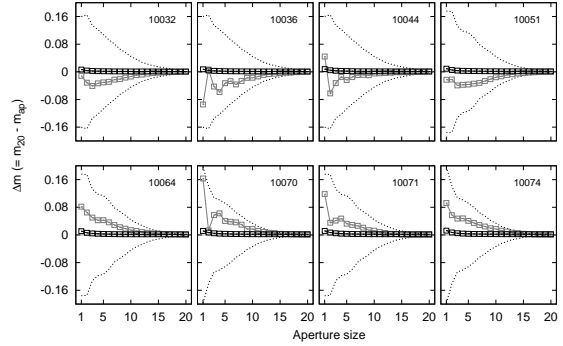}  
    \caption{Systematic variations in the magnitude offsets, $\Delta\mathrm{m}$, for the original multi-aperture magnitude measurements as a function of aperture size, which is marked with a gray dots in all panels.  A positive $\Delta\mathrm{m}$ indicates that the photometric measurements with the corresponding aperture size, $m_\mathrm{ap}$, are brighter than those of reference aperture $m_\mathrm{20}$, while a negative $\Delta\mathrm{m}$ indicates vise versa.  The dashed lines are the rms model profiles introduced in Section 3.2.2.  There is a noticeable distinction between the group 1 (\emph{top panels}) and group 2 (\emph{bottom panels}) when looked at different concentration levels.  But the trends are not fully explained by the difference patterns in $\Delta\mathrm{m}$ because those correlated variations become small after the correction for the PSF variation (black dots).}
    \label{fig:Fig16}
\end{figure}
  
For these two groups, we consider a possible causal relationship between the systematic trends and average object/image properties.  Figure \ref{fig:Fig15} shows the differences in trend, differential magnitudes, sky level, and PSF FWHM between the two groups, respectively.  In the top panel, we plot the magnitude difference in trends, which shows variations in the range of $\pm0.02$ mag.  We suspect that this may be due to the different concentration of star light between these two groups.  It can be checked by using the magnitude difference $\Delta m = m_{20} - m_{ap}$, where $m_\mathrm{20}$ and $m_\mathrm{ap}$ are the reference aperture and the relatively small aperture, respectively.  In fact, we already know that there is a magnitude variation in $\Delta m$ depend on the aperture size due to the field-dependent PSF variation (see Section 3.4).  For example, Figure \ref{fig:Fig16} shows the response of multi-aperture photometry for the two groups of stars at one epoch (MJD = 53726.14817) before and after applying the distortion corrections.  Although the magnitude variation between the group 1 and the group 2 seems to have different behavior as a function of aperture size (gray points), it is negligible after the removal of the PSF variation (black points).  

We also check that the possible contribution of sky level ($\Delta\mathrm{sky}$) and PSF FWHM differences ($\Delta\mathrm{FWHM}$) to the systematic trends on the re-calibrated light curves.  As mentioned by \citet{bra12}, sky over-subtraction may lead to the systematic trends as a function of the PSF FWHM, the amplitude of which increase for fainter stars.  The third and forth panels of Figure \ref{fig:Fig15} show the variation of the mean $\Delta\mathrm{sky}$ and $\Delta\mathrm{FWHM}$, respectively.  In our case, however, the form and amplitude of trends seem independent of sky level and PSF FWHM.  

Some other possible sources that may contribute to the observed systematic trends include: higher order variations in the PSF shape beyond just the FWHM; cross-talk from other amplifiers, or ghosts from bright stars undergoing multiple reflections within the optics; non-uniform variations in the gain; and unmodeled temporal atmospheric variations that are dominated by Rayleigh scattering, molecular absorption by ozone and water vapor, and aerosol scattering \citep{pad08}.  While we find a clear presence of trends that should be removed, we are not able to identify their exact cause.

\subsection{Removal of Temporal Systematics by Photometric De-Trending (PDT)}
In order to reduce systematic effects in photometric time-series data, several methods were introduced (e.g., TFA: \citealt{kov05}; Sys-Rem: \citealt{tam05}; PDT: \citealt{kim09}; CDA: \citealt{mis10}; PDC: \citealt{twi10}).  All of these algorithms share a common advantage that they work without any prior knowledge of the systematic effects.  We use the PDT algorithm, which has been designed to detect and remove spatially localized patterns.  By default, this algorithm works with a set of light curves that contain the same number of data points distributed in the same series of epoch.  In many cases, however, missing data occur when no photometric measurements are available for some stars in a given observed frame.  These missing data can be simply replaced by means, medians, or the values from the interpolation of adjacent data points in each light curve (e.g., \citealt{kov05, kim10}).  Although using the replaced value is the easiest way to reconstruct the light curve to be analyzed, it is not appropriate if the time separation between two subsequent observations is too large.  Instead we use more straightforward approach by applying the PDT algorithm in two separate steps: (i) we construct the master trends from the subset of bright stars, and (ii) de-trend light curves of all stars with most similar master trend and matching time line.  

\begin{figure}[t]
\centering
  \includegraphics[width=\linewidth, angle=0]{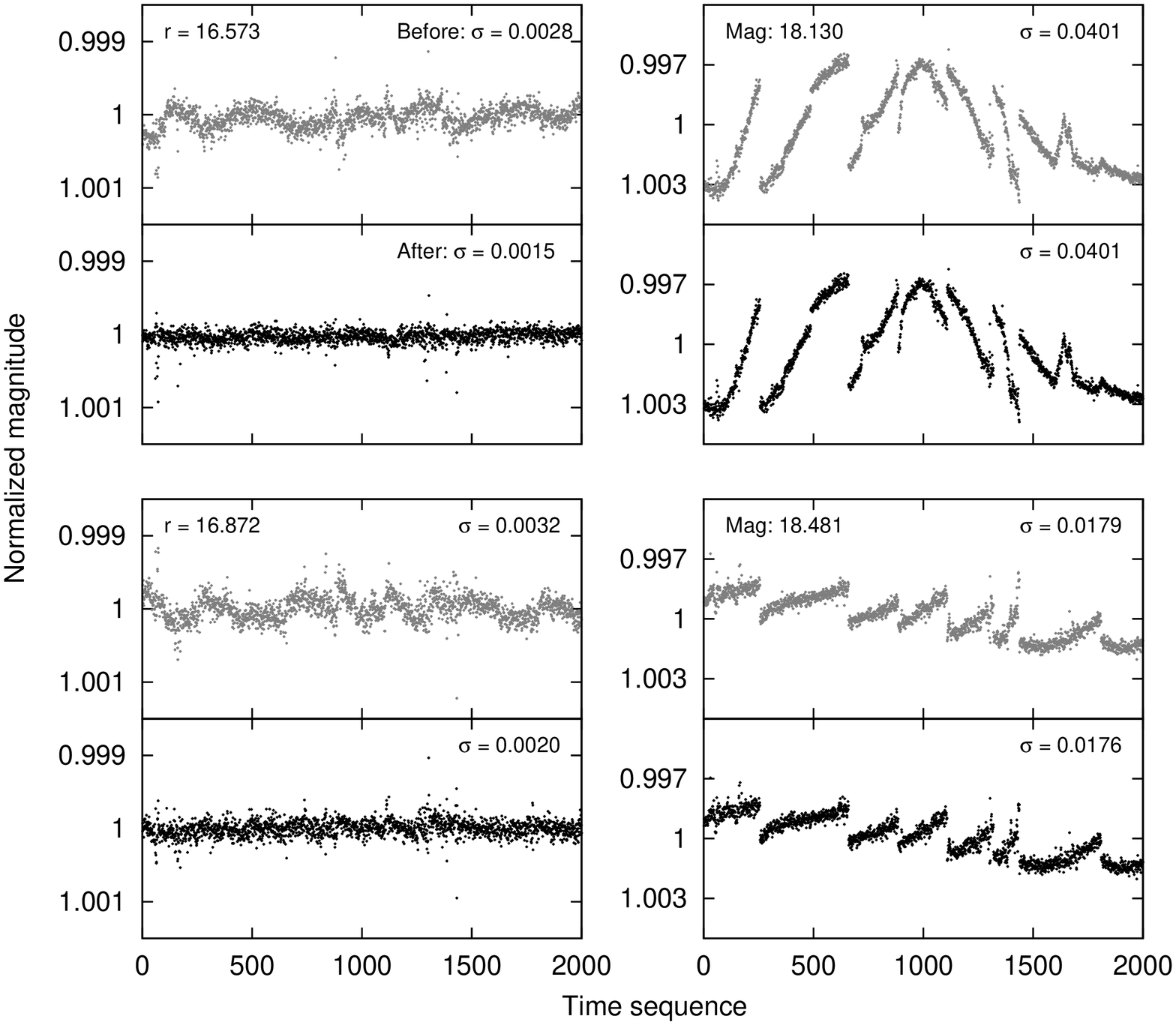}
   \caption{Example of selected light curves before (gray) and after (black) the removal of systematic trends.  The numbers on the $x$-axis indicate the corresponding timestamps between the frame 1 and the frame 2000.  The $y$-axis is $r$-filter magnitude (normalized by its mean value).  While the morphology of the two light curves in the left panels appear to be variable stars of some kind, these turn out to be non-variable after applying the photometric de-trending method.  In the case of the right panels, all true variabilities are preserved from the raw light curves.  From upper left to lower right: ID=10032, ID=10039, ID=170088, and ID=170108.}
   \label{fig:Fig17}
\end{figure}

We briefly describe the main procedure of our de-trending process following the algorithm derived by \citet{kim09}.  We first select the template light curves from bright stars that show the highest correlation in the light-curve features.  The total length of template light curves should be long enough to cover the whole time span of observations.  In this step, we take a sequence of data points, $L_{i}(t_\mathrm{ref})$ as the reference time line.  Using the correlation matrix calculated from Equation (7), we extract all subset of light curves that show spatio-temporally correlated features (i.e., clusters).  Each cluster is determined by hierarchical tree clustering algorithm based on the degree of similarity.  Next, we obtain master trends $T_{c}(t_\mathrm{ref})$ for each cluster by weighted average of the normalized differential light curves, $f_{i}(t_\mathrm{ref})$:
\vspace*{0.1 cm}
\begin{displaymath}
T_{c}(t_\mathrm{ref}) = \frac{\sum^{N_{c}}_\mathrm{i=1}w_{i}f_{i}(t_\mathrm{ref})}{\sum^{N_{c}}_\mathrm{i=1}w_{i}},
\end{displaymath}
\begin{displaymath}
f_{i}(t_\mathrm{ref}) = \frac{L_{i}(t_\mathrm{ref}) - \overline{L}_{i}}{\overline{L}_{i}},
\end{displaymath}
\begin{equation}
w_{i} = \frac{1}{\sigma_{f_{i}}^{2}},
\end{equation} where $N_{c}$ is the total number of light curves in each cluster $c$, $\overline{L}_{i}$ is mean value of \itshape{i}\upshape{th} light curve, and $\sigma_{f_{i}}$ is the standard deviation of $f_{i}(t_\mathrm{ref})$.  After determining the master trends, we de-trend the light curves of all stars with matching master trend and time line.  We adjust the temporal sequence of measurements for the master trends $T_{c}(t_{i})$ by that of individual light curves to be de-trended $L_{i}(t_{i})$.  Because each light curve is assumed as a linear combination of master trends and noise, we can determine the optimal solution by minimizing the residual between the master trends and the light curve:
\begin{displaymath}
\hat{f_\mathrm{i}}(t_{i})=\sum^{m}_{c=1}\beta_{ic}T_{c}(t_{i}) + \epsilon_{i}(t_{i}),
\end{displaymath}
\begin{equation}
\hat{f_{i}}(t_{i})=\frac{L_{i}(t_{i}) - \overline{L}_{i}}{\overline{L}_{i}}.
\end{equation} where $m$ is the total number of master trends and $\beta_{ic}$ are free parameters to be determined by means of minimization of noise term $\sum_{t} \epsilon_\mathrm{i}(t_{i})^{2}$. 

Figure \ref{fig:Fig17} shows examples of our light curves before and after removing the systematic trends.  The algorithm we used for de-trending removes only the systematic variations that are shared by light curves of stars in the adjacent sky regions (left panels), while all kinds of true variabilities are preserved (right panels).

\begin{figure}[t]
\centering
    \includegraphics[width=1.0\linewidth, angle=0]{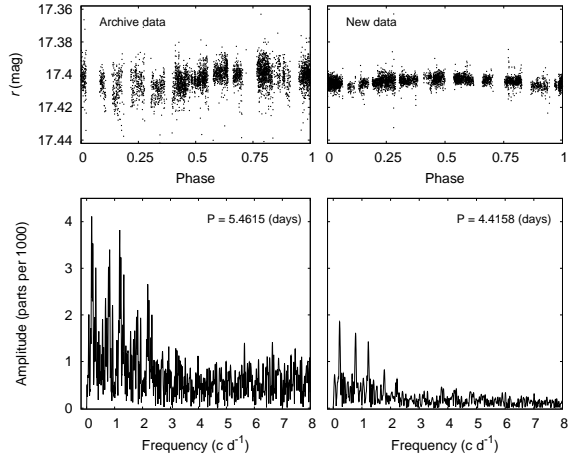}
    \caption{\emph{Top panels}: archival (left) and our final (right) light curves of a periodic variable star, V427.  The light curves are folded by the best-fit period of 5.4615 \citep{har08b} and 4.4158 days, respectively.  Such an period difference for the same star comes from data itself (e.g., both the different noise levels and the different data sampling intervals).  \emph{Bottom panels}: the resulting amplitude spectrum of each light curve is calculated with \texttt{PERIOD04} package.}
    \label{fig:Fig18}
\end{figure}

\begin{figure}[t]
\centering
    \includegraphics[width=1.0\linewidth, angle=0]{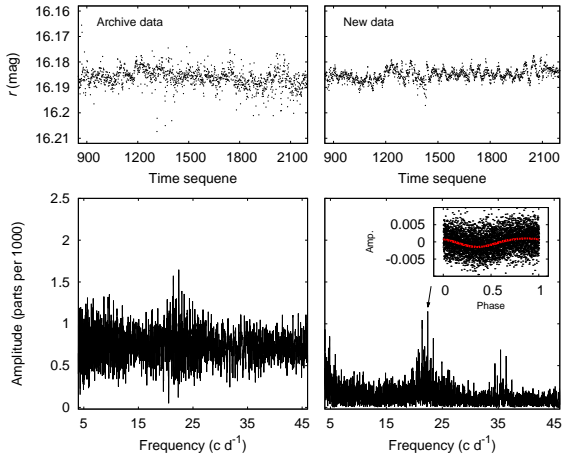}
    \caption{\emph{Top panels}: archival (left) and our final (right) light curves of low-amplitude pulsating variable star, V2276.  The archival data is not adequate to discriminate a signal of astrophysical origin from the noise of the data stream.  \emph{Bottom panels}:  the resulting amplitude spectrum of each light curve is calculated with \texttt{PERIOD04} package.  A significant peak is detected at the frequency 22.3979 days$^{-1}$ (indicated by arrow).  The phased diagram of this candidate frequency is shown in the low-right panel.  We see that the model fits well the overall pulsating variability (red line).}
    \label{fig:Fig19}
\end{figure}

\begin{figure}[t]
\centering
    \includegraphics[width=1.0\linewidth, angle=0]{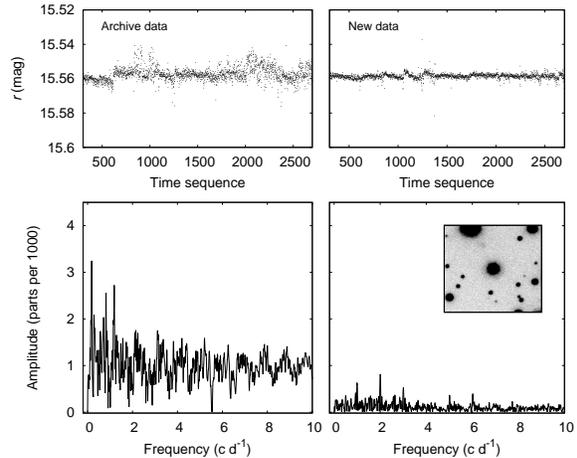}
    \caption{\emph{Top panels}: archival (left) and our final (right) light curves of a star, V347.  This star turns out to be non-variable in the new data.  \emph{Bottom panels}: all other details are same as Figure \ref{fig:Fig19}, but the subfigure of low-right panel is 200$\times$200 thumbnail image of the target star.  There is no potential sources of contamination to hamper the interpretation of the power spectrum.}
    \label{fig:Fig20}
\end{figure}

\section{IMPACT OF THE NEW CALIBRATION OF PERIOD SEARCH}
The usefulness of our photometry is tested for a set of variable stars.  We immediately find abundant cases of improvements in the following three aspects: (i) refinement of the derived period, (ii) detection of a new significant peak in the periodogram, and (iii) separation of non-variable candidates where systematics in the light curves were mistaken for true variability.  For each case, we compare light curves and power spectra for archival data and our data.  

For the first example, we show that a new photometric measurement and calibration allowed us to derive a much improved refinement of the light curves and of the derived periods (Figure \ref{fig:Fig18}).  We performed a Lomb-Scargle (L-S: \citealt{sca82}) search of both archival and new light curves for periodic variable star (V427).  The light curves are folded by the best-fit period of 5.4615\footnote{For the archival data, we adopt the period found on the filtered light curves from \citet{har08b}.} and 4.4158 days, respectively.  We also calculated the false-alarm probability ($\log$ FAP) for each peak and its signal to noise ratio (S/N): $\log$ FAP$_\mathrm{archival}$ = $-28.37$, S/N$_\mathrm{archival}$ = 43.2 for the archival data and $\log$ FAP$_\mathrm{new}$ = $-181.07$, S/N$_\mathrm{new}$ = 84.5 for the new data.  Since we can get better estimation with much lowered minimum FAP value, our new period is the most likely result.  In the bottom panels, the resulting amplitude spectrum was calculated with \texttt{PERIOD04} package.  Since the archival data is more noisy than the new one, it is rather complicated to interpret the peaks of its power spectrum.  

For the second example, we show the newly discovered low-amplitude pulsating variable star (Figure \ref{fig:Fig19}).  We used the \texttt{PERIOD04} package to find multiple pulsation periods.  The whole process of identifying, fitting, and pre-whitening successive frequencies was repeated until no significant frequencies were found.  We adopt a conservative approach in selecting the statistical significant peaks from the amplitude spectrum.  A S/N amplitude ratio of 4.0 is a good criterion for independent frequencies, equivalent to 99.9\% certainty of variability \citep{bre93}.  While no clear periodicity was found in the archive data, our amplitude spectrum shows a clear excess of power centered at 22.3979 days$^{-1}$ with peak amplitudes of about 1 mmag (S/N$_\mathrm{new}$ = 9.02).  
    
The last example is the opposite case of the second.  Figure \ref{fig:Fig20} shows that this object is unlikely to be a variable source because there is no evidence for any significant peaks, which indicates that the variations are mostly noise.  Extensive study on variabilities will be presented in Paper II.

\section{CONCLUSION}
In this paper, we introduce a new time-series photometry with multi-aperture indexing and spatio-temporal de-trending techniques, together with complex corrections to minimize instrumental biases.  We used the archival, high-temporal time-series data from one-month long MMT/Megacam transit survey program.  The re-calibration of the archival data has made several improvements as follows: (i) the photometric information derived from the multi-aperture indexing measurements is useful to obtain the best S/N measurement, but also to diagnose whether or not the targets are contaminated;  (ii) the resulting light curves utilize nearly 100\% of available data and reach precisions down to sub mmag level at the bright magnitude end without the need to throw out many outlier data points, which makes it possible to preserve data points that show intrinsic sudden variations such as flare events;  (iii) corrections for position-dependent PSF variations and de-trending of spatio-temporal systematic trends improve the quality of light curves; and (iv) new photometry enables us to determine the variability nature and period estimate more accurately. 

While this study deals with a particular set of data from MMT, we find our approach has a potential for other wide-field time-series observations.   Multi-aperture indexing measurement is a powerful tool in isolating and even correcting various contaminations.  Spatio-temporal de-trending is also very useful in removing systematics caused by PSF variation and even non-uniform extinction of thin clouds across the FOV.  \citet{cha13} proved this for different sets of archival survey data.

\acknowledgments
This research was supported by Basic Science Research Program of the National Research Foundation of Korea (2011-0030875).   Y.-I.B. is grateful for support from KASI-Yonsei DRC program of Korea Research Council of Fundamental Science and Technology (DRC-12-2-KASI).  We thank the MMT/M37 survey team for the kind provision of raw image data.  Dr. Kim, D.-W. helped us to test the modified photometric de-trending algorithm.  Additionally, we would like to thank the anonymous referee for many helpful comments, including the suggestion to use PSF photometry to further justify the successful removal of systematics (Section 3.5.2).


\clearpage
\end{document}